\documentclass[%
 reprint,
superscriptaddress,
 amsmath,amssymb,
 aps,
floatfix,
]{revtex4-1}

\usepackage{graphicx}
\usepackage{dcolumn}
\usepackage{bm}
\usepackage{hyperref}
\usepackage[mathlines]{lineno}

\newcommand{\ket}[1]{\left| #1 \right>} 
\newcommand{\bra}[1]{\left< #1 \right|} 
 
\begin{document}

\title{Topical Review: Spins and mechanics in diamond}

\author{Donghun Lee}
\thanks{These authors contributed equally to this work}
\affiliation{Department of Physics, University of California Santa Barbara, Santa Barbara, CA 93106}
\affiliation{Department of Physics, Korea University, Anam-dong 5, Seongbuk-gu, Seoul, South Korea}

\author{Kenneth W. Lee}
\thanks{These authors contributed equally to this work}
\affiliation{Department of Physics, University of California Santa Barbara, Santa Barbara, CA 93106}

\author{Jeffrey V. Cady}
\affiliation{Department of Physics, University of California Santa Barbara, Santa Barbara, CA 93106}

\author{Preeti Ovartchaiyapong}
\affiliation{Department of Physics, University of California Santa Barbara, Santa Barbara, CA 93106}

\author{Ania C. Bleszynski Jayich}
\email[email:]{ania@physics.ucsb.edu}
\affiliation{Department of Physics, University of California Santa Barbara, Santa Barbara, CA 93106}

\date{\today}

\begin{abstract}
There has been rapidly growing interest in hybrid quantum devices involving a solid-state spin and a macroscopic mechanical oscillator. Such hybrid devices create exciting opportunities to mediate interactions between disparate qubits and to explore the quantum regime of macroscopic mechanical objects. In particular, a system consisting of the nitrogen-vacancy defect center in diamond coupled to a high quality factor mechanical oscillator is an appealing candidate for such a hybrid quantum device, as it utilizes the highly coherent and versatile spin properties of the defect center.  In this paper, we will review recent experimental progress on diamond-based hybrid quantum devices in which the spin and orbital dynamics of single defects are driven by the motion of a mechanical oscillator. In addition, we discuss prospective applications for this device, including long range, phonon-mediated spin-spin interactions, and phonon cooling in the quantum regime. We conclude the review by evaluating the experimental limitations of current devices and identifying alternative device architectures that may reach the strong coupling regime.
\end{abstract}

\maketitle

\tableofcontents

\section{Introduction}
Advances in quantum technology over the last few decades have led to remarkable control of individual quantum systems, opening new arenas of previously impossible technologies enabled by the strange nature of quantum mechanics \cite{AwschalomReview}. However, practical quantum devices will likely integrate multiple physical systems in such a way that the combined advantages of the total system mitigate the weaknesses of the individual systems \cite{KurizkiPNAS}. For example, a quantum information processor may use solid-state spins as quantum bits (qubits) in a quantum memory register due to their long coherence times, while utilizing superconducting Josephson junctions as computational qubits due to their fast processing capabilities. Realizing these types of hybrid architectures requires a universal bus interface to mediate interactions and information transfer between the various components of the device. Traditionally, single photons have served as the primary interconnect between remote quantum systems, and have been successfully used in some important applications in quantum information science \cite{BlattNaturePhotonics}. However, transferring information between remote quantum systems of disparate energy scales remains challenging, and may benefit from new types of quantum interfaces. 

Mechanical oscillators in the quantum regime offer a promising solution to this challenge because mechanical motion can couple to a wide range of quantum systems through a variety of interactions \cite{HybridMechBook,SklanAIPAdvances,Poot}. For example, mechanical systems can couple to photons through radiation pressure forces, charge qubits through the Coulomb interaction and even spin qubits through magnetic dipole forces (see Fig. 1). Importantly, these coupling mechanisms are conservative, allowing for coherent transfer of quantum information. The ability of a mechanical oscillator to store quantum information is characterized by the quality factor, $Q$, which determines the rate at which the oscillator exchanges energy with the environment. The advent of novel nanofabrication techniques has led to the development of high-$Q$ mechanical resonators over a broad range of frequencies \cite{PreetiAPL,DegenNatComm, BurekAPL,BarclayPRX,Mitchell,DiamondOMCrystals}, establishing hybrid mechanical devices as an attractive, feasible alternative to traditional quantum devices.

The synergistic nature of hybrid mechanical architectures offers several advantages for both qubits and mechanical oscillators. Traditionally, coherent control of qubits is carried out with electromagnetic fields that are delivered through an antenna or waveguide\cite{JelezkoPRL}. However, mechanical degrees of freedom can provide unique and promising ways to probe and control qubits. For instance, as we discuss in section III, mechanically-induced strain can enable direct transitions between spin states of solid-state defects that are forbidden by magnetic dipole selection rules \cite{FuchsPRL,MacQuarrieOptica,Barfuss}. In addition, phonons associated with vibrations of macroscopic mechanical resonators can extend and propagate over long distances while maintaining coherence, and provide a novel mechanism for long-range interactions between qubits \cite{RablNatPhys,Bennett,Habraken,UniversalSAW,GeometricPhase}. 

\begin{figure}[h]
\includegraphics[width=8.6cm]{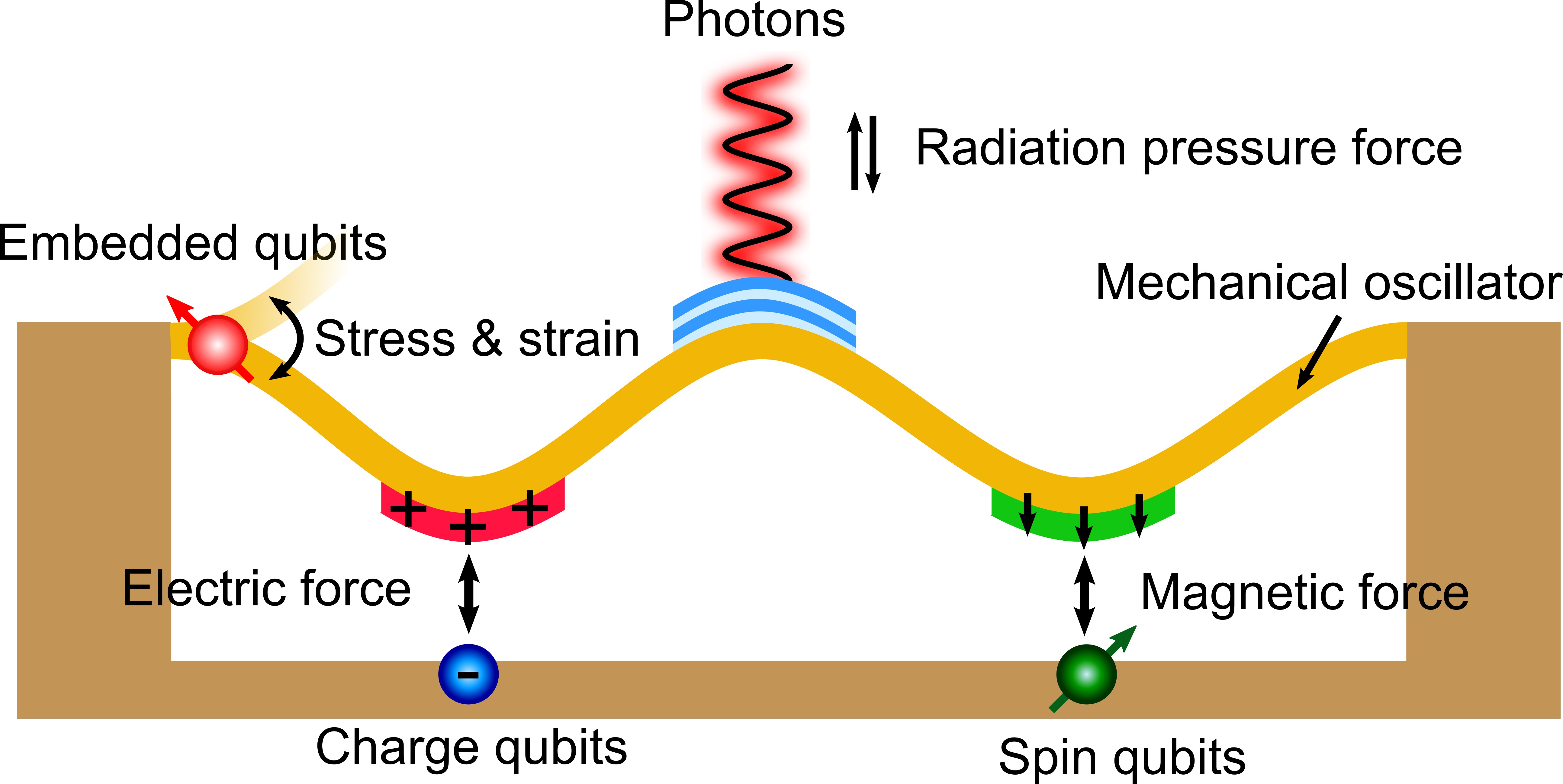}
\caption{Mechanical oscillator as a universal quantum interface between disparate qubits. Here, photons, charge qubits, spin qubits and embedded qubits interact with the mechanical oscillator through radiation pressure, the Coulomb interaction, magnetic dipole forces, and strain respectively.}
\end{figure}

From the perspective of mechanical oscillators, qubits offer a powerful route to generate and detect non-classical mechanical states due to their inherent nonlinearity. Nonlinear interactions allow for preparation of a non-Gaussian mechanical state, such as a Fock state or Schrodinger cat state, from an initial Gaussian state. Realizing non-Gaussian mechanical states remains an important oustanding challenge, as they are suggested to be a necessary requirement for universal quantum computation and expected to be more suitable than Gaussian states for fault-tolerant quantum information processing and secure quantum communication \cite{Andersen,Lund,Takeda}. Moreover, nonlinear interactions can then be used to probe the dynamics of such non-Gaussian states. In the field of cavity optomechanics, nonlinear interactions between cavity photons and the mechanical oscillator can be achieved in the strong, single-photon coupling regime \cite{AspelmeyerReview}. However, realizing strong coupling in cavity optomechanical devices is challenging due to the weak nature of radiation pressure forces, and hence the dominant interaction is typically linear even when the cavity is externally pumped. Using this linear interaction, recent experiments have realized ground state mechanical cooling \cite{TeufelNature,PainterNature,Underwood,PetersonPRL}, photon-phonon entanglement \cite{PalomakiNature,PalomakiScience} and mechanical squeezed states \cite{Wollman,LecocqPRX}, but have been unable to produce non-Gaussian states. A hybrid qubit-mechanical device would naturally provide this non-linear interaction \cite{nonlinearTLS}, and would further benefit from the mature control techniques developed for two-level systems, offering a promising alternative to standard optomechanical devices. 

To date, hybrid mechanical systems have been realized with various two-level systems in order to address issues in a broad range of fields, from fundamental physics to quantum information processing \cite{HybridMechBook}. The two-level systems in these devices include superconducting circuits \cite{SchwabNature,cleland2010,PirkkalainenNature,Etaki}, ultracold atoms \cite{Jockel,CamererPRL,Wang2006,Hunger}, quantum dots \cite{Metcalfe,YeoNatureNano,Montinaro}, and solid-state spins and defects \cite{RugarNature,ArcizetNatPhys,shimonscience,FuchsPRL,NatureComm,Teissier,Barfuss,FuchsPRB,Golter,Kenny,MacQuarrieOptica,HongNanoLetters,Meesala,Golter}. An important figure of merit for such devices is the single phonon cooperativity \cite{RablNatPhys} $C=g^2\gamma^{-1}\Gamma^{-1}$, which compares the single phonon coupling strength between the qubit and mechanical oscillator, $g$, to the intrinsic dissipation of the qubit ($\Gamma$) and mechanical oscillator ($\gamma$). When $C>1$, the interactions between the qubit and mechanical oscillator are coherent, indicating the quantum coupling regime. Table I lists examples of such hybrid systems that have been experimentally demonstrated so far. Interaction mechanisms are indicated with experimentally measured values of $g$, $\Gamma$, $\gamma$, and $C$ with relevant experimental conditions and parameters.

\begin{table*}[h!]
\begin{center}
\begin{tabular}{c}
\includegraphics[width=17.8 cm]{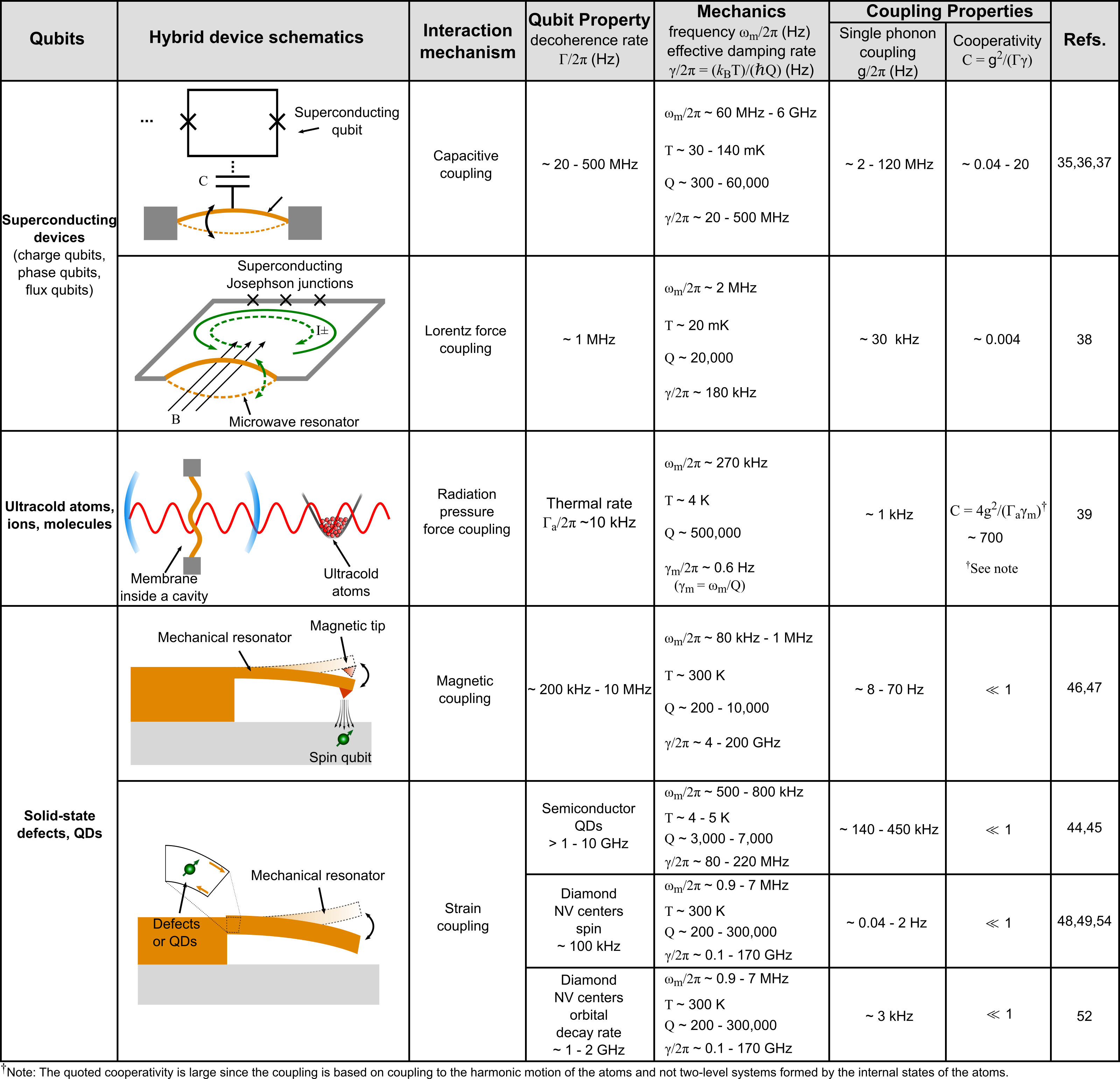}
\end{tabular}
\end{center}
\caption{Examples of experimentally demonstrated hybrid systems combining mechanical oscillators and qubits. The calculated single phonon couplings and cooperativities are calculated from experimental results and not theoretical predictions. We note that the quoted results for ultracold atoms are based on mechanical coupling to the harmonic motion of an ensemble of ultracold atoms as opposed to an ensemble of atomic qubits, and therefore cannot be directly compared to the other listed cooperativities.}
\end{table*}

The field of magnetic resonance force microscopy (MRFM) led to the first experimental realization of a mechanical resonator coupled to a single quantum two-level system \cite{RugarNature}. In ref. \cite{RugarNature}, a magnet-tipped cantilever was used to detect the magnetic force exerted by a single electron spin. Driven oscillations of the magnetic tip were used to periodically invert the spin population, resulting in an alternating magnetic force on the cantilever. This back-action force shifted the mechanical resonance frequency, which was measured with lock-in techniques. The techniques developed in this pioneering work provided a strong foundation for future generations of hybrid mechanical devices. 

The next implementations of hybrid mechanical devices utilized superconducting quantum circuits\cite{SchwabNature,cleland2010,PirkkalainenNature, Etaki}, which are leading candidates for hybrid mechanical devices due to their strong interaction strength with mechanical motion. The motion of mechanical resonators can couple to superconducting qubits via electrostatic \cite{SchwabNature,cleland2010,PirkkalainenNature} or Lorentz force \cite{Etaki} interactions. Interestingly, the quantum regime of coupling has already been demonstrated for both superconducting charge and phase qubits \cite{cleland2010,PirkkalainenNature}. In one such device, strong capacitive coupling to a microwave-frequency mechanical resonator enabled coherent, single-phonon control of mechanical states with a Josephson phase qubit \cite{cleland2010}. 

The next generation of devices coupled mechanical oscillators to even more microscopic systems: ultracold atoms \cite{Jockel,CamererPRL,Wang2006,Hunger} and artificial atoms in the solid state \cite{Metcalfe,YeoNatureNano,Montinaro,ArcizetNatPhys,shimonscience,FuchsPRL,NatureComm,Teissier,Barfuss,FuchsPRB,Golter,Kenny,MacQuarrieOptica,HongNanoLetters,Meesala}.  Hybrid systems based on ultracold atoms benefit significantly from the mature quantum control techniques developed for atomic systems. Thus far, experiments have demonstrated coupling of a cloud of ultracold atoms either directly to a nearby cantilever \cite{Wang2006,Hunger} or indirectly to the drumhead modes of a membrane oscillator via radiation pressure forces \cite{CamererPRL,Jockel}. The quoted cooperativity for atom-based systems in Table I is based on coupling between the mechanical motion of a membrane and the collective harmonic motion of trapped ultracold atoms from ref. \cite{Jockel}. In the future, these systems aim to explore coupling of the internal degrees of freedom of the atoms to the mechanical motion.

Hybrid systems involving quantum dots \cite{Metcalfe,YeoNatureNano,Montinaro} and crystal defects \cite{ArcizetNatPhys,shimonscience,FuchsPRL,NatureComm,Teissier,Barfuss,FuchsPRB,Golter,Kenny,MacQuarrieOptica} retain many of the ideal qualities of ultracold atoms, but reside in a more accessible solid-state platform. In particular, solid-state qubits can be brought into close proximity to mechanical elements and in some cases, couple to the mechanical oscillator in novel ways inaccessible to cold atoms. For example, crystal strain induced by the motion of a mechanical oscillator can couple to the orbital and spin degrees of freedom of embedded quantum dots or crystal defects \cite{Metcalfe,YeoNatureNano,Montinaro,NatureComm,Kenny,FuchsPRL,FuchsPRB,MacQuarrieOptica,Teissier,Barfuss,Meesala}. Technologies incorporating mechanics and solid-state qubits are still in the initial stages of research and development and correspondingly, the cooperativities measured in initial experiments are still small. However, with the continual development of nanofabrication techniques to construct high-Q mechanical oscillators combined with the significant efforts to mitigate decoherence in solid-state qubits, the future for such hybrid mechanical devices looks promising \cite{shimonscience,Kenny,NatureComm}.

In this paper, we will review hybrid mechanical systems based on the nitrogen-vacancy defect center (NV center) in diamond. These devices are promising due to the advantageous properties of the NV center. The NV center is particularly attractive for its highly coherent electron spin, narrow optical transitions, ability to couple to long-lived intrinsic nuclear spins, exquisite spin and orbital control protocols, ability to be controllably placed in a nanoscale volume, and ability to interface with a variety of external degrees of freedom. These advantageous properties make the NV center an excellent candidate for integration in hybrid quantum technologies. In section II, we will give a brief introduction to the NV center and discuss the fabrication and characterization of high Q diamond mechanical oscillators. We will then introduce various implementations of this hybrid device that use either magnetic or strain fields to mediate interactions between the NV center and the mechanical oscillator. In section III, we will discuss novel interactions enabled by hybrid NV-mechanical devices where the mechanical resonator is used to control the spin and orbital dynamics of single NV centers. In section IV, we discuss two important, future applications of this device: single-spin cooling of a mechanical resonator and phonon-mediated spin-spin interactions. In sections V and VI, we will conclude the paper by examining the prospects and challenges for future devices.

\section{Hybrid system based on NV centers in diamond and mechanical resonators}
\subsection{The NV center in diamond}
\begin{figure*}[t]
\includegraphics[width=17.8 cm]{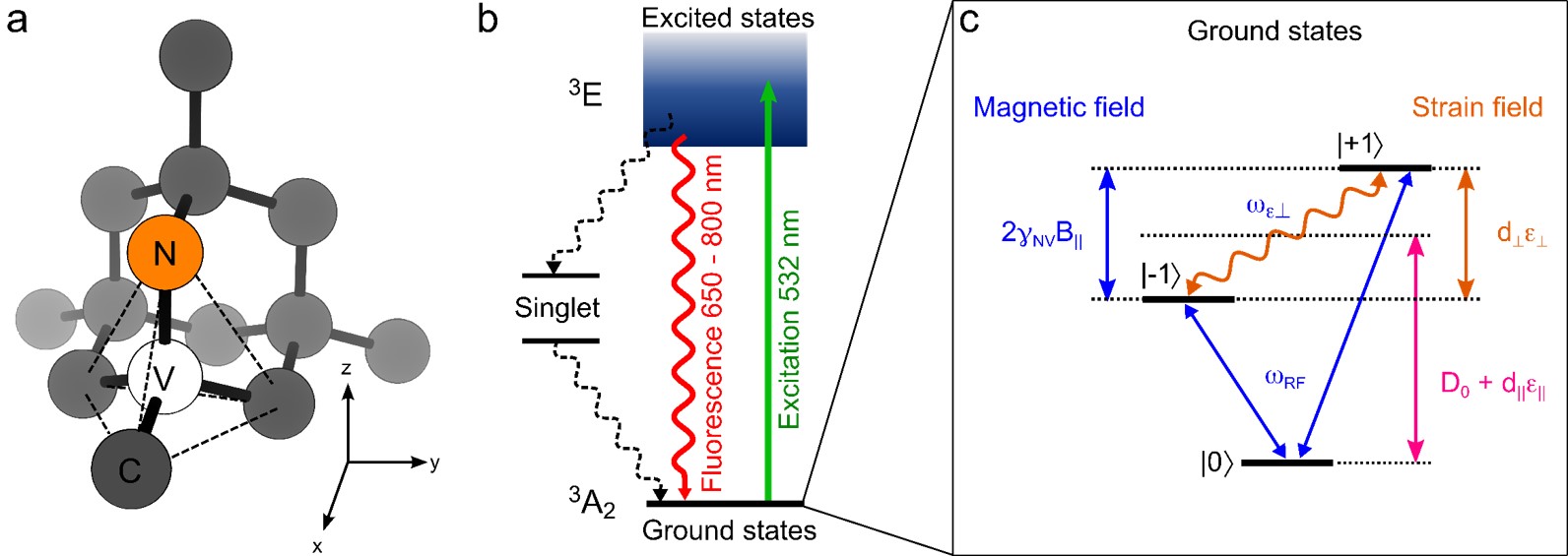}
\caption{Structural, electronic, and optical properties of the diamond NV center. (a) The NV center consists of a substitutional nitrogen atom adjacent to a lattice vacancy and the 3  nearest neighbor carbon atoms to the vacancy. The trigonal structure of the NV is described by the $C_{3v}$ point group symmetry. (b) Simplified electronic structure and optical properties of the NV center at room temperature. Off resonant excitation at 532 nm excites the NV from the $^3A_2$ ground state to the $^3E$ excited state through its absorption phonon sideband in a spin conserving transition. The NV relaxes back to the ground state by emitting a photon into the zero-phonon line ($\sim$ 637 nm) or phonon sideband ($\sim$640-800 nm) or a non-radiative decay through the metastable singlet state. (c) Fine structure of the spin-triplet $S=1$ ground state. External perturbations such as magnetic fields (blue) and crystal strain (orange) alter the spin levels as described in the main text.}
\end{figure*}

The nitrogen-vacancy (NV) center in diamond is an atomic-scale crystal defect that balances the isolation of atomic systems with the strong interactions of solid-state qubits. Its excellent and versatile quantum properties make it promising for use in a wide variety of applications, ranging from quantum information science to nanoscale quantum sensing of biological and condensed matter systems \cite{JelezkoWrachtrup,HansonAwschalom,Schirhagl,LilyPhysToday,RondinReview}. Qubits can be encoded in the orbital or spin degrees of freedom of the NV center. Alternatively, qubits can be encoded in nuclear spins that are coupled to the electron spin of the NV center via hyperfine interaction. The electron spin of the NV center is highly coherent and is regularly used as a qubit. The NV spin can be initialized through optical pumping and optically read out through spin-dependent fluorescence techniques. In addition, the spin states can be coherently and precisely manipulated with microwave magnetic fields. Importantly, the spin of the NV center exhibits long coherence times even at room temperature ($T_2\sim$ ms) \cite{IsotopeEngineering,MaurerScience}, which makes it an appealing and unique system for room temperature quantum information processing. Furthermore, the spin's sensitivity to magnetic\cite{FakeTaylor,MazeNature,WolfPRX}, electric \cite{Dolde,DoldePRL}, strain \cite{NatureComm}, and temperature fields \cite{ToyliTemp,AcostaTemp,NeumannNanoLetters} combined with its atomic-scale resolution make the NV a versatile quantum sensor. 

The remarkable properties of the NV center are in large part afforded by its host material: diamond \cite{WeberPNAS}. First, diamond has a large band gap of 5.5 eV. The electronic bound states of the NV center reside deep within the bandgap, and hence experience negligible coupling to the valence and conduction bands. In addition, diamond's large Debye temperature ($T_D=2200$ K) results in a small thermal population of phonons even under ambient conditions and therefore limits spin-lattice relaxation. Furthermore, advances in chemical vapor deposition (CVD) diamond growth now allow for single-crystal diamond substrates with 99.999$\%$ $^{12}$C isotopic purity \cite{IsotopeEngineering,OhnoAPL}, mitigating decoherence caused by fluctuating $^{13}$C spins in the environment. Finally, diamond has small spin-orbit coupling, and thus undesired spin-flips in electronic transitions are largely avoided. 

Structurally, the NV center is a point defect in the diamond lattice in which a substitutional nitrogen atom is adjacent to a lattice vacancy (Fig. 2a). The electronic structure of the NV center consists of charge, spin, and orbital degrees of freedom. While several charge states of the NV center have been investigated, the negatively charged NV center has the most attractive quantum properties and therefore we will restrict our discussion to the NV$^-$ state, which we henceforth refer to as NV. The electronic ground state of the NV center consists of an orbital-singlet/spin-triplet ($^3A_2$), and has been the subject of most NV related research to date (Fig. 2b).  In the ground state manifold, the $m_s=\pm 1$ spin sublevels are degenerate in energy and split from the $m_s=0$ by $D_0=2.87$ GHz due to spin-spin interactions \cite{Jero,MarcusGroundState}. The $^3A_2$ ground state is connected to an orbital-doublet/spin-triplet excited state $^3E$ by an optical dipole transition that can be excited resonantly ($\lambda_{ZPL}=637$ nm) or off-resonantly through an absorption phonon sideband \cite{Marcus}. At cryogenic temperatures ($T< 20$ K), fine structure emerges in the $^3E$ manifold due to spin-orbit and spin-spin interactions, resulting in six distinct electronic levels: two ($\ket{E_x}$ and $\ket{E_y}$) with $m_s=0$ and four ($\ket{A_1}$, $\ket{A_2}$, $\ket{E_1}$, and $\ket{E_2}$) that are entangled states of non-zero spin and orbital angular momentum \cite{Jero,MarcusNJOP}. At room temperature, electron-phonon interactions within the excited state manifold rapidly mix the orbital levels, averaging to an effective orbital singlet with a fine structure that strongly resembles the ground state, but with $D_{es}=1.42$ GHz \cite{Marcus}. As we discuss in more detail later, the orbital levels of the NV center are highly sensitive to electric fields and crystal strain. Because the NV center has small, non-zero spin-orbit coupling, the spin levels will also be sensitive to electric fields and strain \cite{MarcusGroundState,Dolde,NatureComm}. 

A unique property of the NV center is that its electronic structure allows for all-optical initialization and readout of the spin levels \cite{WeberPNAS,ISCPRL}. The $^3A_2\leftrightarrow ^3E$ transition is a spin-conserving transition for both resonant and non-resonant excitation. However, this optical transition is not a closed transition -- a spin-selective intersystem crossing allows for $m_s=\pm 1$ spin population to be non-radiatively transferred from the $^3E$ manifold to a metastable singlet state, from which population decays back into the $^3A_2$ manifold \cite{ISCPRL} (Fig. 2b). When the NV center is subject to continuous optical excitation, this process allows the spin to be optically pumped into the $m_s=0$ sublevel of the ground state in about 1 $\mu$s. In addition, the finite lifetime and effective non-radiative nature of the metastable state leads to a spin-dependent fluorescence intensity, with the $m_s=\pm 1$ spin levels exhibiting approximately 60$\%$ of the fluorescence intensity of the $m_s=0$ level \cite{Marcus}. This significant difference in the fluorescence allows for optical discrimination of the spin state. At cryogenic temperatures, single-shot readout of the spin state can be accomplished using resonance fluorescence techniques \cite{SingleShot}.

\begin{figure*}[!]
\includegraphics[width=17.8 cm]{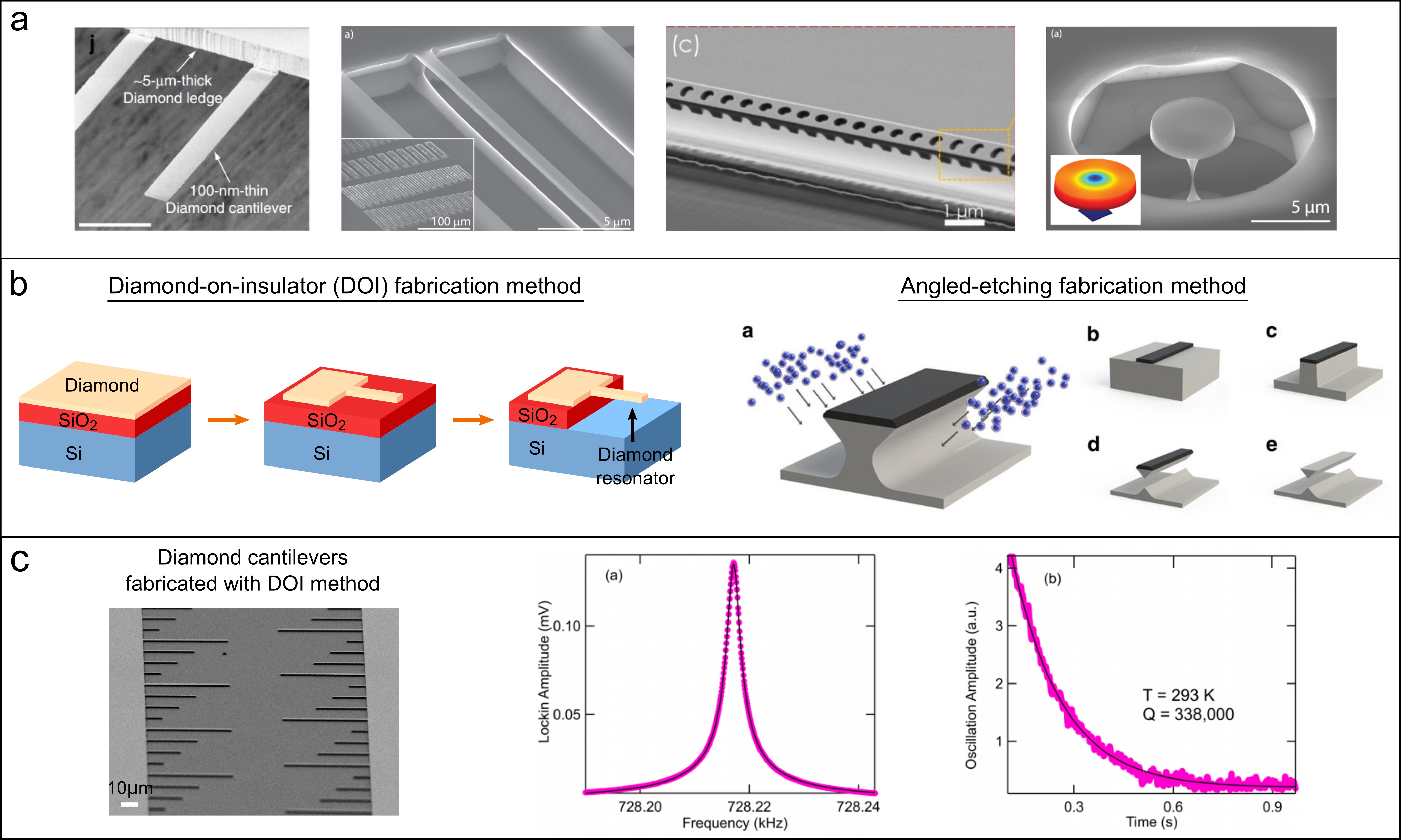}
\caption{Fabrication and characterization of diamond mechanical oscillators. (a) Examples of various single-crystal diamond mechanical structures such as cantilevers \cite{DegenNatComm}, nanobeams \cite{BarclayPRX}, phononic crystals \cite{DiamondOMCrystals}, and nano-disk resonators \cite{Mitchell}. Reprinted with permission from refs. \cite{DegenNatComm,BarclayPRX,DiamondOMCrystals,Mitchell}. Copyright Nature Communications, Physical Review X, Nano Letters (b) Examples of diamond fabrication methods. Left: The ``diamond-on-insulator'' fabrication technique \cite{PreetiAPL}. A diamond thin film is bonded to an SiO$_2$/Si substrate using a low-temperature oxide bonding technique. The mechanical structure is lithographically defined and subsequently formed via etching. The structure is released by etching away the sacrificial oxide layer with hydrofluoric acid. Right: The angled-etching fabrication technique \cite{BurekNatComm}. The mechanical structure is lithographically defined and formed by conventional top-down etching. The sample is then etched at an angle until the structure is released with a triangular cross-section. Images reprinted with permission from \cite{BurekNatComm}. Copyright Nature Communications. (c) Left: Scanning electron micrograph depicting diamond cantilevers fabricated with the DOI approach shown in (b). Center: Mechanical response of one cantilever in the left image shows a resonance at 728 kHz. Right: A mechanical ringdown measurement of this cantilever shows a quality factor $Q>10^5$ at room temperature. Images reprinted with permission from \cite{PreetiAPL}. Copyright Applied Physics Letters}
\end{figure*}

Optical initialization and readout of the NV center allow for optically detected magnetic resonance (ODMR) of the spin, which has been the workhorse for NV-related studies \cite{Gruber}. Individual NV centers can be optically addressed using a confocal microscope. To perform continuous-wave ODMR, the NV is excited with laser light (typically 532 nm) and a microwave magnetic field while the NV fluorescence is monitored with a single photon counting module. When the microwaves are on resonance with a spin transition, the fluorescence of the NV center diminishes as a result of depopulating the $m_s=0$ spin level. One can also perform pulsed ODMR, in which the spin is coherently manipulated with near-resonant microwave pulses and allowed to freely evolve, then optically read out. ODMR spectroscopy has allowed for studies of the response of the NV center to magnetic fields \cite{MazeNature,IsotopeEngineering,MaminScience,Matt,shimonscience,ArcizetNatPhys}, electric fields \cite{Dolde,DoldePRL}, temperature \cite{ToyliTemp,AcostaTemp,NeumannNanoLetters}, and crystal strain \cite{NatureComm,Teissier,Meesala,FuchsPRL}.

\subsection{Diamond mechanical oscillators}

Diamond's excellent mechanical properties make it an attractive platform material for high quality factor mechanical resonators \cite{PreetiAPL,DegenNatComm,DiamondForMechanics}. In addition, the large Young's modulus of diamond ($E=1.2$ TPa) \cite{CVDDiamond} allows for high frequency mechanical resonators to be fabricated with relatively large features, potentially mitigating $Q$ degradation associated with scaling down device dimensions. As we discuss in a later section, this will be an important factor in the success of future diamond optomechanical devices. Importantly, diamond mechanical resonators could be used in an integrated hybrid quantum device where embedded crystal defects can couple to the motion of the resonator through crystal strain.

Fabrication of high-quality diamond mechanical structures thus far has been limited due to challenges in processing and growth. Processing of single-crystal diamond (SCD) is difficult due to diamond's chemically inert and robust nature, and is underdeveloped compared to other technologies such as silicon processing. Moreover, challenges in the growth of high quality single-crystal diamond has limited the availability of suitable substrates. Nonetheless, SCD mechanical oscillators have now been successfully fabricated in many different forms, including clamped beams \cite{PreetiAPL,DegenNatComm,BarclayPRX,BurekAPL}, bulk acoustic resonators \cite{FuchsPRL}, phononic crystals \cite{DiamondOMCrystals}, and nano-disk resonators \cite{BurekNano,Mitchell} (Fig. 3a). To create such structures, two promising processing techniques have emerged, which we briefly discuss in further detail. 

In the first technique, a ``diamond-on-insulator'' (DOI) approach is used \cite{PreetiAPL,DegenNatComm}, in which a thin film of diamond is bonded to an oxide-bearing substrate (Fig. 3b). Due to the prominent lack of commercially available diamond thin films, this technique requires an additional, time-intensive step of producing the thin film, which can be accomplished with ion-implantation techniques \cite{MembranesAdvancedMaterials,MembranesJonathan} or inductively-coupled plasma/reactive ion (ICP/RIE) etching \cite{PreetiAPL}. The mechanical structures are lithographically defined and subsequently created via ICP etching. The sacrificial oxide layer is etched away using a buffered hydrofluoric (HF) acid, which releases the diamond mechanical structure. The second technique involves etching bulk diamond samples in multiple dimensions to release the mechanical structures (Fig. 3b). This process can be accomplished using anisotropic angled etching techniques or with quasi-isotropic ICP/RIE undercut etching \cite{BurekAPL,BurekNano,BarclayPRX}. 

Each method offers distinct advantages in the fabrication and design of diamond mechanical structures. For instance, the DOI approach enables fabrication of two-dimensional structures, which is challenging with the bulk diamond approach. On the other hand, the bulk diamond method is less time-intensive and produces samples that are more compatible with the harsh environments accompanying some post-fabrication processes, such as high temperature annealing. Nonetheless, both methods have been successful in producing high quality factor mechanical resonators. For example, using the diamond-on-insulator approach, diamond cantilevers have been fabricated with quality factors approaching one million at room temperature \cite{PreetiAPL} (Fig. 3c) and even exceeding one million at cryogenic temperatures \cite{DegenNatComm}. Using the bulk diamond technique, 1-D diamond optomechanical crystals were fabricated with a $\sim$ 6 GHz mechanical mode exhibiting a $Q\cdot f$ product exceeding $10^{13}$ \cite{DiamondOMCrystals}. These results demonstrate the feasibility for diamond mechanical structures to provide the basis for hybrid mechanical devices.

Driving the resonant modes of diamond mechanical oscillators has now been accomplished with a variety of methods, including piezoelectric actuation with piezoelectric transducers \cite{PreetiAPL} and interdigitated transducers (IDTs) \cite{Golter}, dielectrophoretic actuation \cite{BurekDriving}, photothermal actuation \cite{BarclayPRX}, and optomechanical actuation \cite{BarclayPRX,DiamondOMCrystals}. Piezoelectric actuation is a simple technique commonly employed in NEMS and MEMS systems, but the capacitance of the piezo stack limits the maximum mechanical actuation frequencies to tens of MHz. On the other hand, IDTs, which also rely on piezoelectric elements, have been used to excite surface acoustic waves in the GHz domain \cite{Golter}, and remain a powerful tool in mechanical technologies. Recently, optomechanical actuation and phonon lasing of GHz frequency mechanical modes was demonstrated in 1-D diamond optomechanical crystals \cite{DiamondOMCrystals}. 

\subsection{Magnetic coupling between spins and mechanics}

The first demonstrations of hybrid mechanical devices based on the NV center exploited the spin sensitivity to magnetic fields. A benefit to using magnetically-coupled spin-mechanical devices is that one can utilize the mature technologies developed in the MRFM community, such as high-$Q$ cantilevers and large magnetic field gradients \cite{MRFMReview}. In a modular device, these critical elements can be optimized and integrated with a pre-selected NV spin that exhibits strong coupling and extended spin coherence times. Magnetically-coupled spin-mechanical devices can be used for a variety of applications in quantum information science, including single-spin cooling of macroscopic mechanical resonators, long-range, deterministic spin-spin interactions and interfacing qubits of various energy scales, such as atoms and nuclear spins \cite{RablNatPhys,RablPRB,RablCoolingHighTemp}. In addition, such a device can be used for nanoscale, magnetic resonance imaging of so-called ``dark'' spins, which may be useful in several applications, such as imaging the dynamics of isolated spin chains or imaging individual spin labels attached to complex biomolecules \cite{GrinoldsNatNano}.

The spin of the NV center is sensitive to magnetic fields through the Zeeman interaction, $H_Z=\gamma_{NV}\mathbf{S}\cdot\mathbf{B}$, where $\gamma_{NV}=2.8$ MHz/G is the NV gyromagnetic ratio and $\mathbf{S}$ is the spin-1 vector operator. A magnetic field applied along the NV symmetry axis breaks the degeneracy of the $m_s=\pm 1$ spin levels, allowing for selective addressing of the $\ket{0}\leftrightarrow\ket{\pm 1}$ transitions with microwaves and formation of a spin qubit. To magnetically couple a mechanical resonator to the spin, the spin must be immersed in a spatially inhomogeneous field such that the position of the spin in the field is controlled by the position of the mechanical oscillator. Suppose that an NV center is located at a position $\mathbf{r_0}$ in a magnetic field $\mathbf{B}(\mathbf{r})$ and the motion of the mechanical oscillator causes the NV center to take small excursions $\delta\mathbf{r}$ from $\mathbf{r_0}$. The Zeeman interaction can then be Taylor expanded to first order in $\delta\mathbf{r}$ such that 

\begin{equation}
H_{Z}\approx \gamma_{NV}S_i\left(B_i(\mathbf{r_0})+\frac{\partial B_i}{\partial r_j}(\mathbf{r_0})\delta r_j\right)
\end{equation}

To date, experiments have focused on the parametric interaction between the spin and the mechanical oscillator, characterized by the interaction Hamiltonian $H_{int}=g(a+a^{\dag})S_z$, where $g=\gamma_{NV}\frac{\partial B_z}{\partial r_j}\delta r_{j0}$ is the single phonon coupling strength, $a$ is the annihilation operator of the mechanical mode, and $\delta r_{j0}$ is the amplitude of  zero point motion of the mechanical oscillator in the $\hat{e}_j$ direction. This parametric spin-phonon Hamiltonian has now been realized in two different architectures. In the work by Arcizet $et\ al.$ \cite{ArcizetNatPhys}, a nanodiamond containing single NV centers was placed on the tip of a SiC nanowire which was then placed in a large magnetic field gradient. As the nanowire oscillates, the NV experiences a spatially-dependent and time-periodic Zeeman shift, thus coupling the motion of the nanowire to the spin. In the work by Kolkowitz $et\ al.$ \cite{shimonscience}, an atomic force microscope (AFM) cantilever with a magnetized tip was positioned over a bulk diamond sample within several nanometers of the target, near-surface NV center (Fig. 4a). As the AFM cantilever oscillates, the NV again experiences a spatially dependent and time-periodic Zeeman shift.

To demonstrate spin-mechanical coupling, Arcizet $et\ al.$ monitored the motion-induced Zeeman shift using continuous-wave ODMR while Kolkowitz $et\ al.$ performed a more sensitive pulsed ODMR technique where the qubit acquires a relative phase due to the motion of the AFM cantilever. In the measurements by Arcizet $et\ al.$, flexural motion of the nanowire is coherently driven with a piezoelectric element while performing continuous-wave electron spin resonance (ESR). The time-varying magnetic field produced by the nanowire motion then frequency modulates the ESR lines, resulting in a broadened, double-peaked spectral lineshape. This lineshape results from averaging over all points of the cantilever's motion, and can be modeled by a Lorentzian whose center frequency is sinusoidally modulated in time.

\begin{figure}[h]
\includegraphics[width=8.6cm]{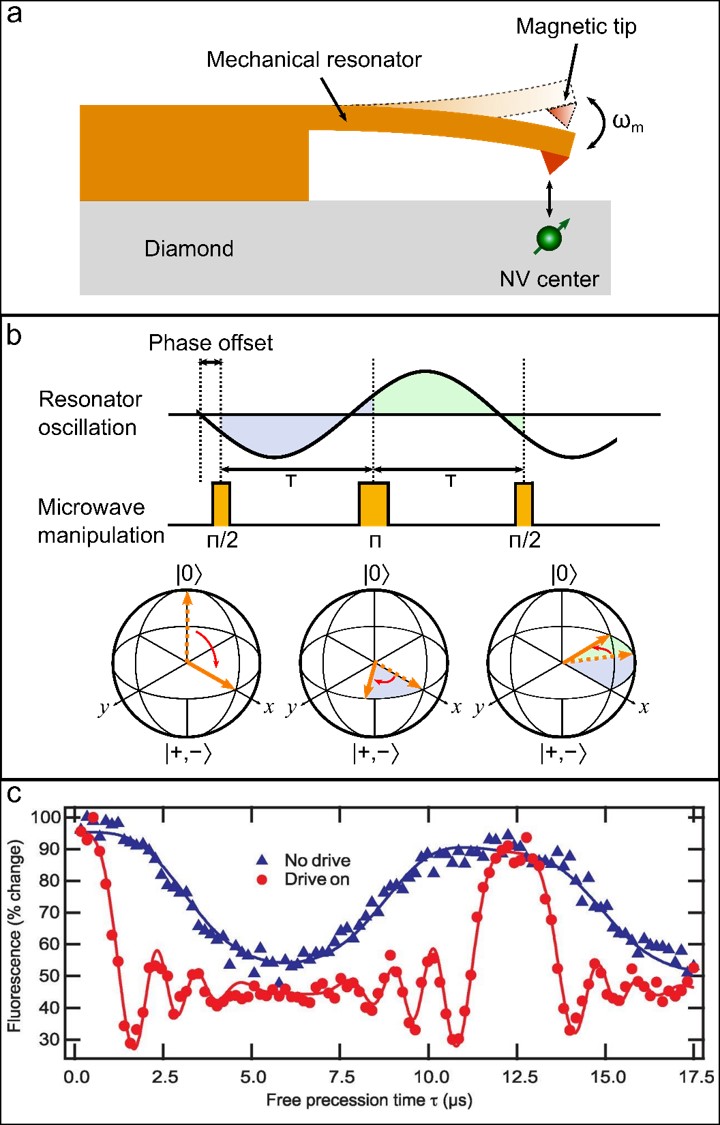}
\caption{Magnetic coupling between spins and mechanics. (a) The mechanical motion of a magnetized AFM tip generates an oscillating magnetic field that couples to the spin states of a nearby NV center. (b) A Hahn echo pulse sequence is synchronized to the resonator's driven motion. The spin-echo pulse allows the mechanically driven relative phase accumulation in the first free evolution period (blue) to add constructively with the relative phase accumulated in the second free evolution period (green). (c) The Hahn echo fluorescence signal as a function of the total free evolution time with (red circles) and without (blue triangles) mechanical driving. With the mechanical drive on, the fluorescence oscillates due to the precession of the spin around the equator of the Bloch sphere. Moreover, strong revivals of the fluorescence occur at each oscillation period of the AFM cantilever. Figure reprinted with permission from \cite{shimonscience}. Copyright Science}
\end{figure}

In the recent experiment by ref. \cite{shimonscience}, the spin of the NV center was used to coherently sense the mechanical motion of the magnetized cantilever. Sensing of the mechanical motion is accomplished with a Hahn echo pulse sequence (Fig. 4b). The sequence begins with a green laser pulse to polarize the NV center into the $m_s=0$  state.  A static magnetic field is applied along the NV symmetry axis to lift the degeneracy of the $m_s=\pm 1$ states and enable selective transitions between the  $m_s = 0$ and $m_s = -1$ (or +1) states. The spin is then prepared in a coherent superposition of the $m_s = 0$ and $m_s = -1$ states with a microwave $\pi/2$ pulse. The oscillating magnetic field generated by the AFM displacement will be then imprinted in the spin evolution as a coherent phase accumulation. In order to constructively accumulate the phase, $\pi$-pulses are applied every half oscillation period to synchronize the pulse to the mechanical motion. The final $\pi/2$ pulse of the Hahn echo converts the accumulated phase to a population difference, which is then read out in the fluorescence signal. Fig. 4c shows the change in the measured fluorescence as a function of the Hahn echo free precession time, $\tau$, when AFM tip is mechanically driven (red). Periodic collapses and revivals of the fluorescence occur at integer multiples of the mechanical oscillation period,  $2\pi/\omega_m=12.5\ \mu$s, reflecting spin evolution due to the driven motion. Under no mechanical drive, the change in the fluorescence is due to hyperfine coupling to proximal, randomly fluctuating $^{13}$C nuclear spins. The overall shape of the fluorescence signal is described by a zero-order Bessel function resulting from a random phase offset between the mechanical motion and microwave pulse train when the experiment is carried out without a phase locked loop (Fig. 4b). 

The Hahn echo sequence described above is inherently sensitive to the variance of the magnetic field produced by the mechanical oscillator. Importantly, this allows for measurement of the Brownian motion of the AFM cantilever, which introduces a noisy magnetic field with a random amplitude and phase but a characteristic frequency of $\omega_m$. To probe the Brownian motion, Kolkowitz $et\ al.$ utilized the higher order dynamical decoupling pulse sequence, XY-4 to extend the coherence time of the NV center and increase the interrogation time, allowing for greater phase accumulation \cite{deLangePRL,deLangeScience}.

The cooperativity of hybrid mechanical devices is determined by the strength of the spin-phonon coupling and the coherence of the spin and mechanical resonator. As we discuss in further detail section V, a few important challenges must be overcome for magnetically coupled hybrid devices to operate in the high cooperativity regime. The limiting factor for such devices is the small single-phonon coupling, with $g=8-70$ Hz the largest demonstrated couplings. This is primarily limited by the strength of the magnetic field gradient. State of the art devices can provide magnetic field gradients up to $10^6$ T/m, and increasing the coupling will require bringing the target NV center closer to the source of the magnetic field gradient. While this is a straightforward task, it will also introduce decoherence channels to the system. For instance, magnetic field noise associated with the field gradient source or thermal drifts of the system can induce decoherence in the spin. Moreover, in the case of the magnetic AFM architecture, bringing the tip in close proximity to the diamond surface can result in non-contact surface friction \cite{SurfaceFriction} which will degrade the $Q$ of the mechanical oscillator. Despite these challenges, it should be possible to operate in the high cooperativity regime by improving system parameters, such as $Q>10^6$ and $T_2>1$ ms, and operating at cryogenic temperatures \cite{RablNatPhys,shimonscience}.

\subsection{Strain-mediated coupling between spins and mechanics}

A hybrid mechanical device that uses crystal strain to interface a mechanical oscillator with an embedded NV center is an attractive, monolithic platform that should address some challenges facing magnetically coupled devices. Strain coupling is intrinsic to the device, and requires no additional external components, such as magnetic tips or cavities. This in principle eliminates drifts in the coupling strength due to thermal drifts that are present in devices where precise alignment of external components to the target quantum system is required. In addition, strain coupling does not generate noisy, stray fields such as Johnson noise, which introduce decoherence to the spin. Moreover, monolithic architectures are more pragmatic in terms of device scalability. As these devices are scaled to host large numbers of qubits and mechanical oscillators, it will be crucial to minimize the number of independent components. However, strain-coupled spin-mechanical devices suffer from a few important drawbacks. First, the spin-strain interaction is inherently weak and correspondingly, all demonstrated strain-coupled devices have smaller zero-point motion couplings than magnetically coupled devices. In addition, while a monolithic architecture eliminates drifts, it also eliminates modularity in the device that is available to magnetically coupled devices in processing and functionality. Finally, without the capability of scanning the position of the NV with respect to the resonator, strain-coupled devices require precise positioning of the spin inside of the mechanical structure. However, recently demonstrated nanoscale NV placement techniques that maintain high NV spin coherence times alleviate the severity of this challenge\cite{Claire}.

From an applications standpoint, hybrid spin-mechanical devices using a crystal strain interface support several unique and important functionalities relevant to the fields of quantum information science and quantum-assisted sensing. Crystal strain associated with phonons of high-$Q$ mechanical resonators or phononic waveguides can be used as an interconnect between remote qubits of disparate energy scales \cite{Habraken,UniversalSAW}. The crystal strain interaction can be used, for example, to prepare multiple NV spins in entangled states, perform deterministic quantum teleportation between NV centers, or generate squeezed spin states for quantum sensing \cite{Bennett,RablNatPhys}. Alternatively, this device could provide a quantum interface between photons of disparate energy scales \cite{Joerg,AndrewsPhotonConversion}. The crystal strain interaction between the NV center and a mechanical resonator also offers a route to cooling the mechanical resonator to its quantum ground state \cite{WilsonRae,Kepesidis}, which would enable fundamental studies of quantum mechanics in macroscopic objects. 

\begin{figure*}[!]
\includegraphics[width=17.8cm]{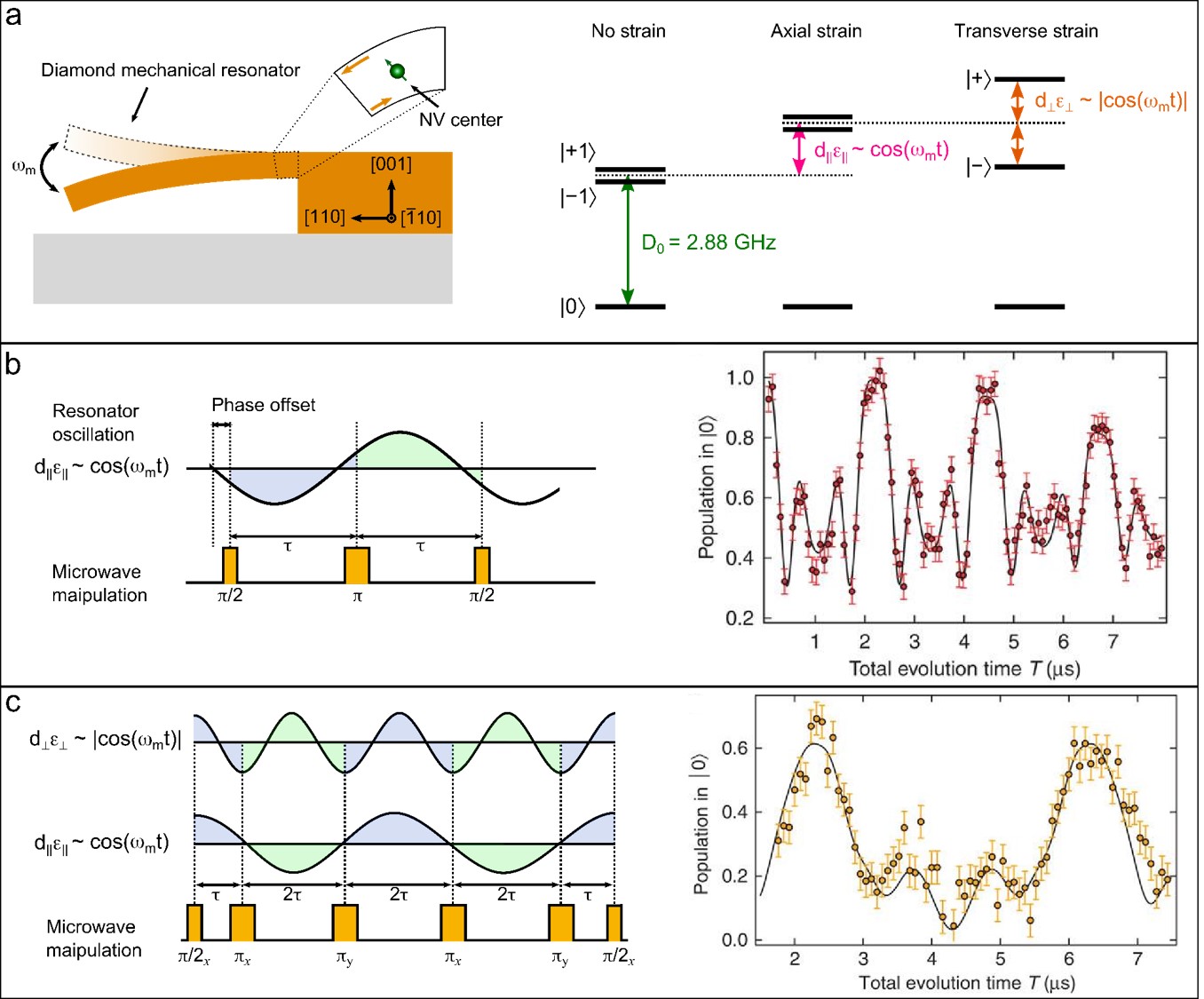}
\caption{Strain coupling between spins and mechanics. (a) The flexural motion of a cantilever deforms the diamond lattice which couples to the ground state spin of an embedded NV center. Axial strain shifts the $\ket{\pm 1}$ spin levels uniformly with respect to $\ket{0}$ while transverse strain splits and mixes the $\ket{\pm 1}$ levels resulting in new eigenstates, $\ket{\pm}$. (b) Left: A Hahn echo measurement is used to extract the axial strain coupling as described in Fig. 4. A magnetic field applied along the NV axis suppresses contributions due to transverse strain. Right: The Hahn echo fluorescence signal demonstrates coupling of the spin to the driven mechanical motion. Data figure reprinted with permission from \cite{NatureComm}. Copyright Nature Communications. (c) An XY-4 pulse sequence is used to enhance and measure the spin sensitivity to transverse strain. Transverse strain and axial strain modulate the spin states at different characteristic frequencies, allowing them to be distinguishable in the spin evolution and fluorescence signal. Data figure reprinted with permission from \cite{NatureComm}. Copyright Nature Communications.}
\end{figure*}

In strain-coupled NV spin-mechanical devices, the vibration of the mechanical resonator deforms the diamond lattice, introducing strain to embedded NV centers (Fig. 5a). The strain interaction with the spin can be modeled as an effective electric field interaction \cite{NatureComm,Dolde,vanOort}. Based on this model, the spin-strain Hamiltonian in the $^3A_2$ ground state manifold with a non-zero magnetic field can be written as, 

\begin{eqnarray}
H_{NV}=&&(D_0+d_{\parallel}\epsilon_{\parallel})S_z^2+\gamma_{NV}\vec{S}\cdot\vec{B}\nonumber\\
&-&\frac{d_{\perp}\epsilon_{\perp}}{2}(e^{i\phi_s}S_+^2+e^{-i\phi}S_-^2)
\end{eqnarray}

We define $S_{\pm}$ as the spin-1 raising and lowering operators $(S_{\pm}=S_x\pm iS_y)$ and $d_{\parallel}$ and $d_{\perp}$ as strain coupling constants to strain parallel ($\epsilon_{\parallel}$) and perpendicular ($\epsilon_{\perp}$) to the NV axis where $\epsilon_{\perp}=\sqrt{\epsilon_{x}^2+\epsilon_{y}^2}$ and $\tan{\phi_s}=\epsilon_y/\epsilon_x$. Here, $\epsilon_{i=\{x,y,z\}}$ refers to the diagonal component of the strain tensor in the NV coordinate basis. 
These effects can be qualitatively understood by the symmetry of the NV structure. As seen in Fig. 2a, the trigonal structure of the NV center belongs to $C_{3v}$ group. The NV preserves its symmetry upon rotation by $2\pi/3$ and $4\pi/3$ around the NV axis and reflection upon three vertical mirror planes containing the NV axis and each carbon atom. Tensile or compressive strain along the NV axis changes the relative distance between the lattice ions but maintains the $C_{3v}$ symmetry. Therefore strain components parallel to the NV axis (axial strain) increase (upon compressive strain) or decrease (upon tensile strain) the energy of the $m_s=\pm 1$ spin states relative to the $m_s = 0$ state (Fig. 5a). On the other hand, strain components perpendicular to the NV axis (transverse strain) break the symmetry, resulting in splitting and mixing of the $m_s=\pm 1$ spin levels \cite{MarcusGroundState}. In the presence of the axial and transverse strain, the eigenvalues of the mixed states relative to $m_s=0$ state can be written as,

 \begin{equation}
 E_{\pm}=D_0+d_{\parallel}\epsilon_{\parallel}\pm\sqrt{(\gamma_{NV}B_z)^2+(d_{\perp}\epsilon_{\perp})^2}
 \end{equation} 
 
The strain coupling constants, $d_{\parallel}$ and  $d_{\perp}$, have been measured in recent experiments \cite{NatureComm,Teissier}, and we will discuss the results from ref. \cite{NatureComm}. As seen in Fig. 5a, a singly-clamped diamond cantilever is strain-coupled to embedded NV centers near the diamond surface. The fundamental flexural mode of the cantilever is resonantly driven at $\omega_m/2\pi \sim 800$ kHz by a piezoelectric actuator. The cantilever motion induces time-varying strain along the cantilever axis ([110] crystal axis) that is proportional to the beam displacement, which can be quantized in terms of the phonon creation and annihilation operators. 

\begin{equation}
\epsilon=\epsilon_0x=\epsilon_0(a+a^{\dag})
\end{equation}

Here, $\epsilon_0$ is the strain induced by the zero-point motion of the cantilever, and $x$ is the displacement of the cantilever normalized to the zero-point motion. Due to the Poisson effect, non-zero strain is also induced along directions perpendicular to the cantilever axis. As a result, the NV center experiences time-varying axial and transverse strains whose relative strengths depend on the NV crystallographic orientation with respect to the cantilever axis. For instance, the NV centers oriented $[\bar{1}\bar{1}\bar{1}]$ or $[11\bar{1}]$ experience predominantly axial strain (cf. transverse strain for NVs along $[\bar{1}11]$ or $[1\bar{1}1]$). Moreover, the presence of a magnetic field ensures that the effects of axial and transverse strain occur at different characteristic frequencies, allowing one to experimentally distinguish the effects. This can be seen in eq. 2, in which motion-induced axial strain,  $\epsilon_{\parallel}$, modulates the qubit energy with a time dependence given by $ \cos{⁡(\omega_m t)}$ while  transverse strain, $\epsilon_{\perp}$ obeys $\sim |\cos{(\omega_m t)}|$.

The axial strain constant was first obtained with the NVs that experience predominantly axial strain. A Hahn echo measurement similar to the one described in the previous section was used to measure the dynamic phase imprinted in the spin evolution resulting from the axial strain. A relatively strong magnetic field of 22 G was applied along the NV axis to suppress the effects of transverse strain during the measurement. The obtained Hahn echo signal evolves at the qubit frequency of $\omega_m$ reflecting the mechanical driving of the spin evolution. From fits to the expected Hahn echo signal for several NV centers, the axial strain coupling constant was obtained to be $d_{\parallel}=13.4\pm0.8$ GHz \cite{NatureComm}. 

For the transverse strain measurement, a weaker magnetic field of $B_z=16$ G was applied to the NVs oriented perpendicular to the cantilever axis and the higher order dynamical decoupling (DD) pulse sequence XY-4 was used. In this case, both the axial and transverse strain affect the qubit population, and beat notes between the transverse and axial strain are written onto the XY-4 signal. From a fit to the expected XY-4 signal, the transverse strain coupling constant was obtained to be $d_{\perp}=21.5\pm1.2$ GHz \cite{NatureComm}. We note that the measurements from ref. \cite{Teissier} obtained similar results for the coupling constants. The experiments in ref. \cite{Teissier} utilized continuous-wave ESR to probe the spectral response of the spin to static bending of a cantilever. Ref. \cite{Teissier} also performed ESR while resonantly driving the cantilever, and observed resolved mechanical sidebands due to axial strain coupling.

\subsection{Strain mediated coupling between orbits and mechanics}

In contrast to the spin states, the orbital states of the NV center are more directly sensitive to crystal strain and can therefore couple more strongly to mechanical motion \cite{Jero,Kenny,MarcusNJOP}. Strain-orbit coupling directly affects the electric dipole transitions of the NV center in both frequency and polarization. Importantly, strain-orbit coupling can then be used to integrate the NV center into hybrid systems consisting of both phonons and photons \cite{Kenny,Kepesidis,UniversalSAW,Habraken,Golter}. The strain-orbit interaction can be carefully controlled with mechanical resonators, allowing for precise, deterministic control of the optical properties of the NV center. This functionality could be useful for photonic applications that require generation of indistinguishable photons, such as entanglement of distant spins\cite{BernienEntanglement,ToganNature,LoopholeFree}. Importantly, the large orbital sensitivity to mechanical motion enables applications that are difficult to achieve using spin-strain coupling, such as phonon cooling of a mechanical resonator and phonon routing\cite{Kepesidis,UniversalSAW,Habraken}. 
However, strain-orbit coupled systems have important drawbacks. The coherence times of the orbital states are intrinsically short, as they are limited by the excited state lifetime ($\sim$ 100 ns \cite{Marcus}), hindering their use as a qubit. In addition, the strong sensitivity of the orbital levels to electric fields and crystal strain results in significant spectral diffusion of the zero-phonon line. As we discuss in section V, this effect is more pronounced for near-surface NV centers and NV centers in processed structures, which are critical to the success of many strain-coupled devices. Nonetheless, the advantages of strain-orbit coupling can still be used to improve devices that use spin-mechanical couplings. For instance, a device may use strain-orbit coupled NV centers to cool a phonon mode that is interacting with a set of NV center spins. Alternatively, an effective spin-phonon interaction can be engineered from the strain-orbit interaction using stimulated Raman transitions, as we discuss in section IV.

The strain-orbit Hamiltonian arises from a modification of the Coulomb interaction between the crystal ions and the electrons of the NV center due to small, relative displacements of the crystal ions. In the limit of small ionic displacements, the leading-order perturbation to the NV Hamiltonian is linear in crystal strain, and can be written as \cite{Jero,Kenny,MarcusNJOP}.

\begin{equation}
H_{strain}=\sum_{i,j}V_{ij}\epsilon_{ij}
\end{equation}

Here $i,j$ sum over cartesian indices defined in the NV coordinate basis, $\mathbf{V}$ is an orbital tensor operator, and $\mathbf{\epsilon}$ is the elastic strain tensor, with both written in terms of the native coordinate system of the NV center as described in the previous section. The effects of crystal strain on the orbital degree of freedom can be elucidated by projecting the strain Hamiltonian onto the irreducible representations of the $C_{3v}$ group, which reflect the symmetry of the orbital wavefunctions.

\begin{equation}
H_{strain}=\sum_{\Gamma}V_{\Gamma}\epsilon_{\Gamma}
\end{equation}

Here, $\Gamma$ indexes the $C_{3v}$ irreducible representations $\{A_1,A_2,E\}$. The components of $\mathbf{V}$ and $\mathbf{\epsilon}$ transform as the quadratic basis functions of the $C_{3v}$ group. For instance, $\epsilon_{zz}$ transforms as $z^2$, and thus transforms as $A_1$. As a result, no components of the strain tensor transform as the $A_2$ irreducible representation.

 Experiments demonstrating strain-orbit coupling have focused on the $^3E$ manifold, which is highly sensitive to strain. The strain-orbit interaction in this manifold can be written as
 
 \begin{equation}
 H_{strain}=H_{A_1}+H_{E_1}+H_{E_2}
 \end{equation}
 
 where
 
\begin{equation}
 H_{A_1}=\left[\lambda_{A_1}\epsilon_{zz}+\lambda_{A_1'}(\epsilon_{xx}+\epsilon_{yy})\right]\mathbf{I}
\end{equation}
 
 \begin{equation}
 \begin{split}
 H_{E_1}&=\left[\lambda_{E}(\epsilon_{yy}-\epsilon_{xx})+\lambda_{E'}(\epsilon_{xz}+\epsilon_{zx})\right](\ket{E_x}\bra{E_x}\\
 &-\ket{E_y}\bra{E_y}+(\ket{E_1}\bra{A_1}-\ket{E_2}\bra{A_2}+\text{h.c.}))
 \end{split}
 \end{equation}
 
  \begin{equation}
 \begin{split}
 H_{E_2}&=\left[\lambda_{E}(\epsilon_{xy}+\epsilon_{yx})+\lambda_{E'}(\epsilon_{yz}+\epsilon_{zy})\right](\ket{E_x}\bra{E_y}\\
 &+i\ket{E_2}\bra{A_1}-i\ket{E_1}\bra{A_2}+\text{h.c.})
 \end{split}
 \end{equation}
 
where $\mathbf{I}$ is the identity operator and $\lambda_{A_1}=-1.95\pm0.29$ PHz, $\lambda_{A_1'}=2.16\pm0.32$ PHz, $\lambda_{E}=-0.85\pm0.13$ PHz, and $\lambda_{E'}=0.02\pm0.01$ PHz are the orbital strain coupling constants \cite{Kenny,DaviesHamer}. Strain of $A_1$ symmetry causes a uniform shift of the $^3E$ manifold in energy, while strain of $E$ symmetry split and mix the fine structure states. Although the $^3E$  manifold is host to 6 total states, only $\ket{E_x}$ and $\ket{E_y}$ have been experimentally investigated for strain-orbit coupling. The $\ket{A_1}$, $\ket{E_1}$, and $\ket{E_2}$ states are not ideal for several strain-orbit coupling applications due to their strong coupling with the metastable singlet state \cite{ISCPRL}. We note that the $\ket{A_2}$ state is isolated from the singlet states and can be used in future strain-orbit coupling applications, as we discuss in section IV. We will henceforth limit our discussion mostly to the $\ket{E_x}$ and $\ket{E_y}$ states.
 
 \begin{figure}[h]
\includegraphics[width=8.6cm]{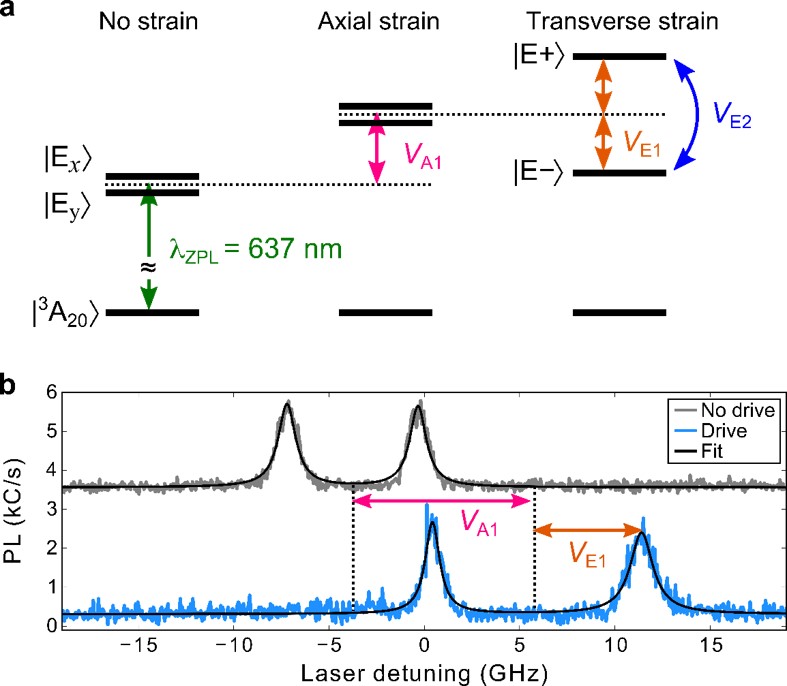}
\caption{Strain coupling between orbits and mechanics. (a) Simplified level structure showing the orbital ground state state $\ket{A}$ and excited states $\ket{E_x}$ and $\ket{E_y}$ connected by an optical zero-phonon line at $\lambda_{ZPL}=637$ nm. Strain of $A_1$ symmetry uniformly shifts $\ket{E_x}$ and $\ket{E_y}$ with respect to the ground state while strain of $E$ symmetry splits $(V_{E_1})$ and mixes $(V_{E_2})$ $\ket{E_x}$ and $\ket{E_y}$. (b) Resonant excitation spectroscopy reveals the $\ket{A}\rightarrow\ket{E_y}$ and $\ket{A}\rightarrow\ket{E_x}$ transition peaks (gray). With mechanical driving (blue), the peaks shift together due to $A_1$ strain and split due to $E$ strain. Black curves are fits to a Lorentzian lineshape}
\end{figure}
 
In recent experiments, crystal strain has been used to interface the orbital levels of the NV center with diamond cantilevers \cite{Kenny} and surface acoustic waves \cite{Golter}. As described in the previous section, crystal strain can be quantized in terms of the phonon creation and annihilation operators of the mechanical mode of interest, leading to the overall orbit-phonon coupling Hamiltonian. 

\begin{equation}
\begin{split}
H&=\Big[g_{A_1}(\ket{E_x}\bra{E_x}+\ket{E_y}\bra{E_y})+g_{E_1}(\ket{E_x}\bra{E_x}\\
&-\ket{E_y}\bra{E_y})+g_{E_2}(\ket{E_x}\bra{E_y}+\ket{E_y}\bra{E_x})\Big](a+a^{\dag})
\end{split}
\end{equation}

where $g_{\Gamma}$ is the single phonon coupling strength for phonons with symmetry $\Gamma$. Crystal strain of $A_1$ and $E_1$ symmetry results in a parametric interaction between mechanical resonator and the $\ket{E_x}$ and $\ket{E_y}$ levels, whereas strain of $E_2$ symmetry directly couples and mixes $\ket{E_x}$ and $\ket{E_y}$ (Fig. 6). Introducing a controlled strain field to the NV center allows for deterministic control of both the frequency and polarization dependence of the zero phonon line \cite{Kenny}, as we discuss in the following section. In addition, the parametric strain-orbit interaction can be used for optomechanical control of the orbital states when the phonon frequency is much larger than the linewidth of the zero-phonon line (i.e. the resolved sideband regime), as was recently demonstrated using surface acoustic waves \cite{Golter}.

\section{Mechanical control of NV centers}

In this section, we review previous experiments demonstrating mechanical control of the orbital and spin degrees of freedom of the NV center. In the first subsection, we discuss how mechanical motion can be used to coherently manipulate the spin of the NV center through magnetic or strain couplings. We follow up this discussion by showing how coherent mechanical driving of the spin can be used to protect the spin from magnetic field noise, which is the predominant source of decoherence for the NV center. We conclude the section by reviewing two complimentary experiments demonstrating mechanical control of the orbital states of the NV center.

\subsection{Coherent mechanical manipulation of the NV spin}

Conventionally, the ground state spin levels of the NV center are manipulated with microwave magnetic fields delivered through an antenna or waveguide near the sample \cite{JelezkoPRL}. In recent years, the use of mechanical oscillators to coherently manipulate and control individual quantum systems has emerged as an attractive alternative approach. In particular, because mechanical degrees of freedom can influence a wide variety of physical systems, they can be used to manipulate several disparate quantum systems. In the case of the NV center, mechanical resonators have now been successfully used to coherently control the spin degree of freedom through magnetic and strain couplings.

The first demonstration of mechanical control of the NV center spin was performed by Hong $et\ al.$ \cite{HongNanoLetters}. In the experiments, a magnetized atomic force microscopy (AFM) tip is brought in proximity to a single NV center near the surface of a bulk diamond sample. When the AFM tip is at rest, the static field produced by the tip splits the $m_s=\pm 1$ spin levels, allowing for selective addressing of the $|0\rangle\leftrightarrow|1\rangle$ transition. Under a resonant mechanical drive, the transverse motion of the AFM tip generates an oscillatory magnetic field that modulates the spin resonance frequencies of the NV center. In order to control both the spin population and phase, an additional microwave field was applied to the NV center and the motion of the AFM tip was synchronized to the optical and microwave control. To control the spin population, the microwave frequency was set to be at the center of the frequency modulation range. As the tip moves, the spin transition is moved through resonance with the applied microwave field, allowing for an adiabatic inversion of the spin population. For each oscillation period, the spin population is inverted twice. To control the spin phase, the spin is prepared in a equal superposition of the spin levels with an RF $\pi/2$ pulse. The motion of the tip imprints a relative phase onto the spin which is determined by the free evolution time of the spin and the oscillation amplitude of the magnetic tip, which can be carefully controlled. To read out the phase of the spin, a second microwave pulse can be applied to convert the relative phase into a population difference which can then be read out optically. 

This approach to coherent mechanical control of the NV spin was then taken a step further in the experiment described in ref. \cite{PhononMollow}, where the authors synchronized the spin dynamics of a single NV center with the motion of a mechanical oscillator, generating a phononic Mollow triplet. In this work, a nanodiamond containing single NV centers was placed on the end of a SiC nanowire which was immersed in a large magnetic field gradient. The oscillation of the nanowire then generates a position-dependent Zeeman shift for the NV center. The authors then applied a microwave field to the NV center, creating microwave-dressed spin states. In this dressed state basis, the parametric interaction between the spin and the mechanical oscillator appears as a transverse interaction, allowing for resonant spin flips. However, this resonance condition requires that the microwave Rabi frequency be equal to the mechanical oscillation frequency. When this resonance condition was met, the spin underwent long-lasting Rabi oscillations exhibiting beatnotes in the spin evolution corresponding to a Mollow triplet structure, a strong signature of spin-locking to the mechanical motion.

While mechanical control via magnetic coupling has opened the door to various applications in quantum information and sensing, it can only generate interactions between the $|0\rangle$ and $|\pm 1\rangle$ spin states. Magnetic dipole selection rules allow for a total change in the spin angular momentum of $\Delta m=\pm 1$. Coherent transitions between the $|\pm 1\rangle$ spin levels are thus difficult to access using conventional pulsed magnetic resonance techniques. Accessing the dipole forbidden $|-1\rangle\leftrightarrow|1\rangle$ transition would open the door to several interesting applications. For instance, direct access to the $|-1\rangle\leftrightarrow|1\rangle$ transition could enable new types of spin-based sensing or even enhance current magnetometry protocols \cite{FangQuantumBeats,MaminDoubleQuantum}. Moreover, combining this with standard microwave spin manipulation allows for simultaneous, coherent addressing of all 3 spin transitions in the ground state of the NV center, forming a closed $\Delta$ system \cite{ShahriarDeltaSystem,KosachiovDeltaSystem}. Such a system would exhibit unconventional spin dynamics that are sensitive to the relative phases of the driving fields, which may find particular value in quantum-assisted sensing and quantum optomechanics.

\begin{figure}[!]
\includegraphics[width=8.6 cm]{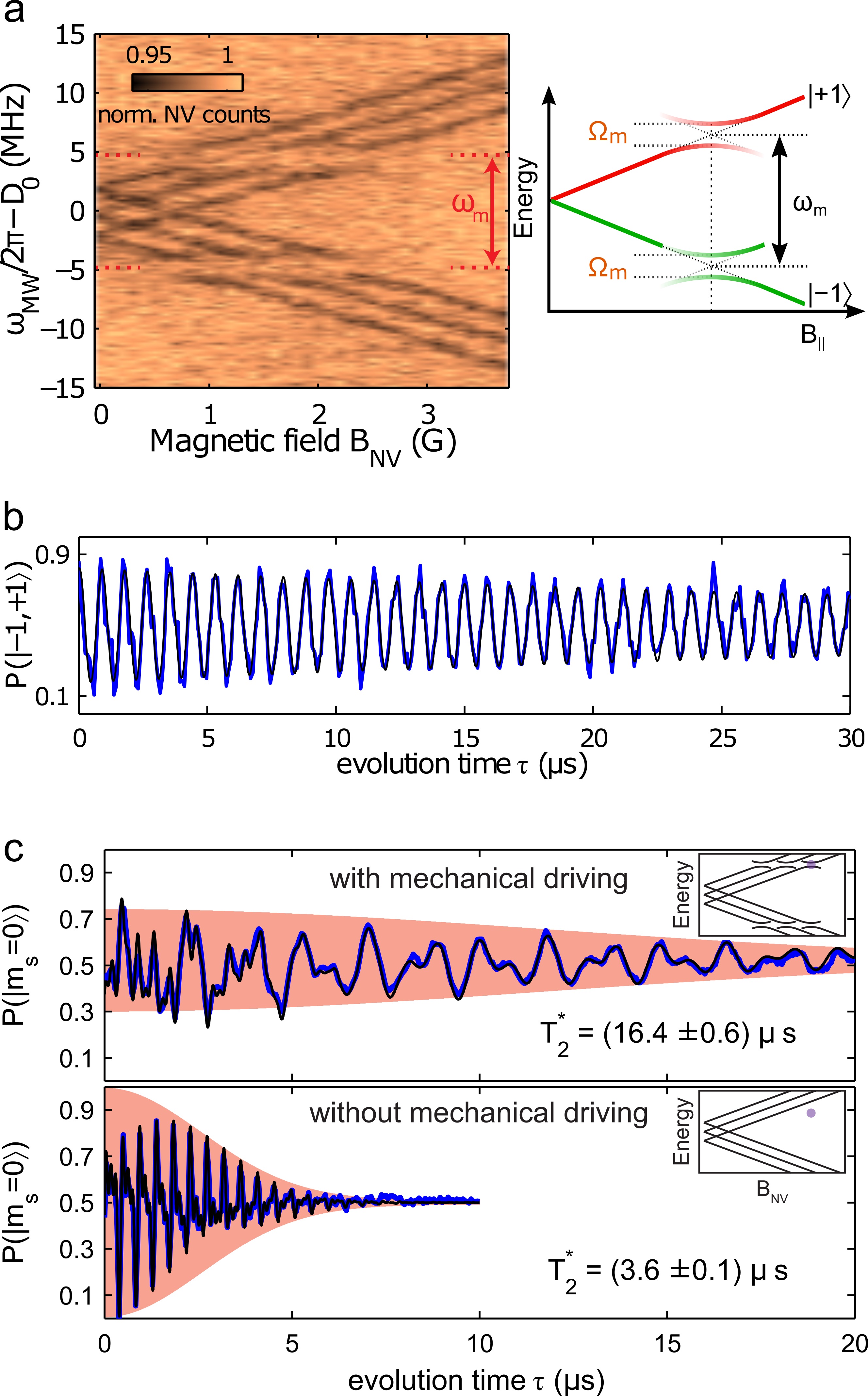}
\caption{Mechanical spin control and decoherence protection. (a) Experimental and theoretical spectra of the $\ket{\pm 1}$ spin levels as a function of the magnetic field applied along the NV axis ($B_{\parallel}$) and with mechanical driving. Each spin level is further split by hyperfine interaction with $^{14}$N, resulting in hyperfine states $\ket{\pm 1,m_I}$. When the $\ket{\pm 1,m_I}$ splitting matches the mechanical oscillation frequency, an Autler-Townes splitting is observed, demonstrating direct coupling of the $\ket{\pm 1,m_I}$ states and formation of dressed-spin states. In the weak driving regime, this splitting is linear in the mechanical Rabi frequency $\Omega_m$. In the strong coupling regime, the linear relationship breaks down. (b) Mechanically driven Rabi flopping between the $\ket{\pm 1}$ states mediated by transverse strain. Reprinted with permission from ref. \cite{Barfuss}. Copyright Nature Physics. (c) The spin coherence time of the NV is increased by approximately an order of magnitude with mechanically driven continuous dynamical decoupling as measured with a Ramsey experiment. The strain-dressed spin states are magnetic field insensitive to first order, as shown at the avoided crossing in (a). Reprinted with permission from ref. \cite{Barfuss}. Copyright Nature Physics}
\end{figure}

Strain-driven spin transitions have now been observed in two different hybrid mechanical architectures. In refs. \cite{FuchsPRL,MacQuarrieOptica} , a high-overtone bulk acoustic resonator (HBAR) was fabricated on a diamond substrate hosting single or an ensemble of spins, and used to generate an acoustic standing wave and hence a time and space-periodic strain field in the diamond. In the work by ref. \cite{Barfuss}, the fundamental mode of a single-crystal diamond cantilever was mechanically actuated with a piezoelectric transducer, introducing an AC strain field to embedded NV centers. Spin transitions between the $|\pm 1\rangle$ spin levels occur when the frequency of the mechanical strain field, $\omega_m/2\pi$, matches the $|\pm 1\rangle$ frequency splitting, $\Delta_B=2\gamma_{NV}B_z$, which is set by an adjustable static magnetic field oriented along the symmetry axis of the NV center. In both architectures, coherent spin manipulation was demonstrated in both the time domain with Rabi oscillations as shown in Fig. 7b and the frequency domain with microwave spectroscopy.

In ref. \cite{Barfuss}, coherent manipulation of the spin was demonstrated in the strong coupling regime and characterized with dressed-state spectroscopy. In particular, Barfuss $et\  al.$ performed electron spin resonance (ESR) measurements of the NV in the presence of mechanical driving and for a variable $\Delta_B$ (Fig. 7a). The microwave probe field was sufficiently low in power to resolve the triplet $^{14}$N hyperfine structure, which splits the $\ket{\pm 1}$ states by $A_{\parallel}=2.1$ MHz into states $\ket{m_s,m_I}$. Including hyperfine interaction, the strain-resonance condition occurs for when the $\ket{-1,m_I}\leftrightarrow\ket{1,m_I}$ splitting is matched to the mechanical frequency, $\Delta_B=\omega_m/2\pi+m_IA_{\parallel}$. When the resonance condition was met, an Autler-Townes splitting was observed in the spectrum, characteristic of strain-dressed spin states. To demonstrate strong driving, Barfuss $et\ al.$ then performed ESR for $\Delta_B=\omega_m/2\pi$ with a variable mechanical Rabi frequency $\Omega_m$. For weak driving where $\Omega_m\ll \omega_m$, the Autler-Townes splitting was linear in $\Omega_m$. However, when $\Omega_m\geq\omega_m$, the linear relationship broke down, demonstrating the onset of the strong-driving regime and the breakdown of the rotating wave approximation. 

In ref. \cite{MacQuarrieOptica}, the first demonstration of a spin coherence measurement in the $\ket{\pm 1}$ basis without double-quantum magnetic pulses was performed. Using mechanical strain pulses, MacQuarrie $et\ al.$ performed a Ramsey experiment on an ensemble of NV centers embedded in an HBAR resonator system. The inhomogeneous spin coherence time $T_2^{\ast}$ of the $\ket{\pm 1}$ qubit was shown to be a factor of two smaller than qubits that utilized the $\ket{0}$ state. The results are in good agreement with theoretical predictions and previous measurements with double-quantum magnetic pulses. Importantly, these results establish that a $\Delta$-system is possible within the NV spin ground state.

\subsection{Enhancing spin coherence with mechanically-driven continuous dynamical decoupling}

Solid-state spins are promising for use as quantum sensors and quantum bits due their long quantum coherence times. For quantum sensing, the inhomogeneous dephasing time, $T_2^{\ast}$, sets the minimum sensitivity of the spin or spin ensemble, whereas in quantum information processing, $T_2^{\ast}$ sets the upper bound on the number of operations that can be performed on a qubit and the storage time of a quantum memory. Conventionally, pulsed dynamical decoupling (DD) has been used to suppress inhomogeneous dephasing and extend the spin coherence time to the homogeneous spin dephasing time $T_2$. This technique has been used to great success for generation of long-lived quantum memories and precision metrology \cite{LilyScience,deLangeScience,deLangePRL,NaydenovPRB}. However, pulsed DD suffers from a few important drawbacks. First, pulsed DD decouples the spin from slowly-varying noise while enhancing the spin's sensitivity in a narrow frequency band. From a sensing standpoint, this eliminates the ability to measure quasi-static fields and significantly limits the bandwidth of the sensor. Second, constructing quantum gates that commute with pulsed DD sequences is technically demanding and requires large overhead, limiting the speed and fidelity of qubit operations \cite{DynamicalDecoupling}. 

Very recently, continuous dynamical decoupling (CDD), has emerged as an alternative, advantageous approach to suppressing spin decoherence \cite{CDDPRL,FuchsPRB,Barfuss}. In this technique, the spin is continously and coherently addressed with an external control field, creating new ``dressed'' spin states which are less sensitive to environmental fluctuations and exhibit an extended $T_2^{\ast}$ spin dephasing time. For the electron spin of the NV center, dephasing is primarily caused by magnetic field noise from nearby paramagnetic spins, which couples to the NV spin through the Zeeman Hamiltonian $H_Z=\gamma_{NV}B_zS_z$ \cite{deLangeScience,BryanPRL}. Experiments on the NV center typically utilize one of the $|0\rangle\leftrightarrow|\pm 1\rangle$ spin transitions, which are magnetic field sensitive to first order. As described in the previous section, an AC transverse strain can be used to coherently drive transitions between the $|\pm 1\rangle$ spin levels. The resulting strain-dressed spin levels, $|\pm\rangle$, are linear combinations of $|\pm 1\rangle$ and satisfy $\langle\pm|H_Z|\pm\rangle=0$. A qubit using any of the new spin eigenstates would therefore be magnetic field insensitive to first order and exhibit significantly longer coherence times.

In the refs.\cite{FuchsPRB,Barfuss}, strain-driven CDD was used to significantly enhance the spin coherence times of single NV centers embedded inside mechanical resonators. To demonstrate this effect, both groups performed Ramsey interferometry on a single spin with and without mechanical driving. In the absence of mechanical driving, the bare spin states of the NV exhibited short coherence times of a few microseconds. However, with resonant mechanical driving, $T_2^{\ast}$ increased due to the strain-dressing of the spin states. As the mechanical driving strength increased, the observed spin coherence time increased and eventually saturated, marking an increase of $T_2^{\ast}$ by an order of magnitude.

While the $\{\ket{0},\ket{+}\}$ and $\{\ket{0},\ket{-}\}$ qubits are insensitive to magnetic field noise, they are still vulnerable to fluctuations in temperature, electric fields, and strain. In the work of MacQuarrie $et\ al.$, thermal instabilities on the order of $0.25^{\circ}$C limited the enhanced spin coherence time to $T_2^{\ast}=12\ \mu$s. To mitigate thermal instabilities, both refs. \cite{FuchsPRB,Barfuss} utilized a qubit spanned by the $|\pm\rangle$ spin levels, which are insensitive to both magnetic and thermal fluctuations. In the work of Barfuss $et\ al.$, the spin coherence was improved from 3.6 $\mu$s to a saturating value of 16.4 $\mu$s. It was suggested that due to the proximity of the NV centers to the diamond surface, electric field noise dominates residual dephasing. In addition, the saturation of $T_2^{\ast}$ with mechanical driving strength was attributed to technical noise in the mechanical driving field. In the work of MacQuarrie $et\ al.$, utilizing the $\ket{\pm}$ subspace improved the spin coherence time to $T_2^{\ast}= 15.4\ \mu$s, which was estimated to be limited by technical noise in the driving field. However, MacQuarrie $et\ al.$ demonstrated further improvement in the spin coherence by isolating a single hyperfine level associated with a nearby $^{13}$C spin. Selective strain-dressing maximally protects the target hyperfine level at the expense of the other hyperfine level. Isolation of a single hyperfine level led to a 19-fold enhancement of the spin coherence time in the $\{\ket{\pm}\}$ basis, with $T_2^{\ast}>50\ \mu$s. 

\subsection{Mechanical control of the optical properties of a quantum emitter}

A common way to couple remote, matter qubits via a photonic channel is to interfere and detect single photons emitted by the qubits. Quantum interference requires the photons to be indistinguishable. As a quantum emitter in the solid state, the NV center faces the challenge of overcoming spectral inhomogeneities resulting from locally varying electrostatic and strain environments in the host diamond \cite{BernienEntanglement,ToganNature}. The conventional solution to this challenge utilizes multi-axis DC electric fields that are delivered via metal electrodes placed nearby or on the diamond sample \cite{BassettPRL,AcostaPRL}. While this technique has found success, it introduces additional channels for spectral diffusion through electric field noise produced by the electrodes or changes in the electrostatic environment. In ref. \cite{Kenny}, an alternative approach was taken in which mechanically-induced strain was used to deterministically tune both the frequency and polarization dependence of the zero-phonon line of two spatially separated NV centers. 

The device in ref. \cite{Kenny} consists of a single-crystal diamond cantilever hosting near-surface NV centers. The flexural motion of the cantilever produces a time-varying crystal strain which couples strongly to the orbital degree of freedom of the NV center and modulates both the frequency and polarization dependence of the zero-phonon line. In particular, ref. \cite{Kenny} investigated the electronic states with spin projection $m_s=0$, consisting of two excited state levels ($\ket{E_x},\ket{E_y}$) and one ground state level ($\ket{A}$), which are consistently used in photonic applications involving the NV center \cite{BernienEntanglement,SingleShot,BernienPRL,BuckleyScience}. Fig. 8a shows the excitation spectrum of two distinct NV centers: one located in the cantilever (NV I) and one located in the diamond bulk far away from the cantilever (NV II). In the absence of mechanical driving, their $E_x$ and $E_y$ transitions differ by several optical linewidths. To tune the $E_x$ and $E_y$ transitions of NV I into resonance with those of NV II, Lee $et\ al.$ resonantly drove the cantilever at varying deflection amplitudes and synchronized the excitation measurement to the mechanical motion. Specifically, Lee $et\ al.$ performed stroboscopic excitation measurements with respect to the cantilever motion which isolated the response of the NV center for a well-defined cantilever position. 

 \begin{figure}[!]
\includegraphics[width=8.6cm]{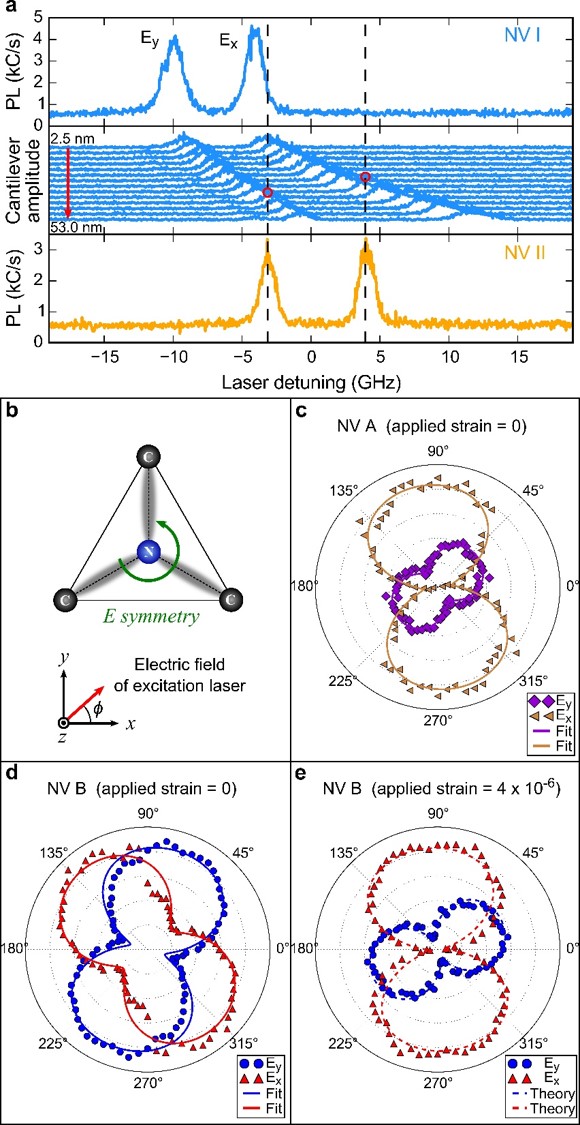}
\caption{Mechanical strain control of the optical properties the NV center. (a) Resonant excitation spectroscopy of two spatially separated NV centers shows the $E_y$ and $E_x$ transitions for each NV differ in frequency by several GHz. NV I is located inside of a cantilever and NV II is located in the sample bulk. With mechanical driving, the transitions of NV I are carefully tuned to those of NV II (marked by red circles). Figure reprinted with permission from \cite{Kenny}. (b) Strain of $E$ symmetry rotates the orientation of the $E_x$ and $E_y$ transition dipole moments. A schematic defining the linear polarization angle is shown with respect to the principal axes of the NV center. (c) The dipole excitation pattern for the $E_x$ and $E_y$ transitions of NV A, which is located in the sample bulk. (d) The dipole excitation pattern for NV B which is located in the cantilever. In the absence of driving, the dipole excitation pattern does not match that of NV A. (e) With mechanical driving, the dipole orientations of NV B are matched to those of NV A. Figures adopted with permission from \cite{Kenny}.}
\end{figure}

Fig. 8a shows stroboscopic excitation measurements for various cantilever deflection amplitudes. At amplitudes of 43 nm and 30 nm respectively, the $E_x$ and $E_y$ transition frequencies of NV I match those of NV II. By applying feedback to the stroboscopic circuit, this technique could be extended to reduce low frequency contributions to spectral diffusion, such as local charge fluctuations \cite{AcostaPRL}. Moreover, this technique could be useful when scaling the number of NV qubits in a diamond chip. A device containing an array of cantilevers with different and spectrally resolved resonances can address and tune a high density of NV qubits without requiring individual electrical contacts and leads for each qubit.

The polarization dependence of the optical transitions of spatially separated NV centers can also be matched using this mechanical control scheme. Fig. 8c-e shows the polarization dependence of the $E_x$ and $E_y$ transitions for two different NV centers: NV A which is located inside the sample bulk and NV B which is embedded in a cantilever. Due to their inhomogeneous environments, their optical transitions exhibit different polarization dependences. In Fig. 8e, the polarization dependences of the $E_x$ and $E_y$ transitions of NV B were matched to those of NV A by driving the cantilever at a particular deflection amplitude and measuring stroboscopically. The dashed curves in Fig. 8e correspond to the expected polarization dependences, taking into account the cantilever amplitude and distortions in the shape of the dipole excitation patterns caused by non-zero ellipticity in the laser beam and saturation of the optical transitions. The excellent agreement between measured and expected polarization dependences highlights the deterministic nature of this mechanical control scheme. 

The device in ref. \cite{Kenny} does not allow for simultaneous and independent control of $E_1$ and $E_2$ strain, and thus the frequency and polarization dependence of the zero-phonon line cannot be simultaneously and independently controlled \cite{BassettPRL}. Universal control of the zero-phonon line could be enabled by driving two mechanical modes that generate different strain profiles, which can be easily accomplished by driving a single piezoelectric element at two frequencies. Photon-mediated interactions can also be carried out using the current device by frequency matching and externally manipulating the polarization state of the emitted photons.

\subsection{Optomechanical control of the NV orbital states}

The use of phonons for both quantum control and generation of entanglement has found greatest success in the trapped ion community. For instance, laser-assisted spin-dependent forces can be used to prepare the internal states of the ions in entangled states via virtual excitations of the vibrational modes of the ion chain \cite{CiracZoller,MolmerSorensen,MonroeGate}. In a recent work by Golter $et\ al.$ \cite{Golter}, the authors took a first step toward realizing a solid-state analog to the trapped ion system, in which a combination of lasers and phonons were used for quantum control of a single NV center. Specifically, surface acoustic waves (SAWs) were used to perform coherent, phonon-assisted optical transitions on single NV centers in the resolved sideband regime.

SAWs, such as Rayleigh waves, propagate along the surface of any elastic material and extend about one acoustic wavelength below the surface \cite{UniversalSAW}. In ref. \cite{Golter}, SAWs are generated in a 400 nm piezoelectric ZnO layer sputtered on the diamond surface that is externally driven with aluminum interdigitated transducers (IDTs). The phonon frequency was determined by spacing of the IDT electrodes and the speed of sound of ZnO, and found to be $\omega_m/2\pi=900$ MHz. 

The SAWs generated by the IDTs introduce a time-periodic strain field to NV centers that couples to the orbital degree of freedom of the NV center. To demonstrate phonon-assisted optical transitions, Golter $et\ al.$ focused on the parametric strain-orbit interaction, with the interaction Hamiltonian

\begin{equation}
H_{int}=g(a+a^{\dag})\ket{E_y}\bra{E_y}
\end{equation}

where $g$ is the single-phonon coupling and $a$ is the annihilation operator of the SAW mode. This parametric interaction introduces phonon sidebands on the $E_y$ optical transition. In the resolved sideband regime where the phonon frequency greatly exceeds the linewidth of the electronic transition, the sidebands can be selectively addressed. In the work described here, The NV centers addressed were naturally occurring in the diamond substrate and exhibited relatively narrow optical linewidths of approximately 175 MHz, allowing for operation in the resolved sideband regime. In the Lamb-Dicke regime where the overall strain-orbit coupling is much smaller than the phonon frequency, the effective Hamiltonian governing the first red sideband transition is given by ($\hbar=1$)

\begin{equation}
H_{rsb}=\frac{\Omega_0g}{2\omega_m}\left(\sigma^+a+\sigma^-a^{\dag}\right)
\end{equation}

where $\Omega_0$ is the natural Rabi frequency and $\sigma^{\pm}$ are the raising and lowering operator for the two-level optical transition. In the experiment, the SAWs were mechanically driven, and hence the phonon mode can be described as a large amplitude coherent state with amplitude $\alpha$. Therefore, the Rabi frequency for the red sideband transition is given by $\Omega_{rsb}=\Omega_0g\alpha/\omega_m$. 

Golter $et\ al.$ demonstrated phonon-assisted transitions in both the frequency and time domain. A 532 nm laser was used to initialize the NV center into the $m_s=0$ ground state spin level. The $\ket{0}\leftrightarrow\ket{E_y}$ transition is resonantly addressed with a laser near 637 nm that is gated with an AOM. The population in the ground and excited state manifold is read out through the photoluminescence of the NV center. 

The phonon sidebands were probed in the frequency domain using resonant excitation spectroscopy. In the absence of mechanical driving, a single Lorenztian peak appeared in the spectrum corresponding to the $\ket{0}\rightarrow\ket{E_y}$ transition. With mechanical driving, sidebands appeared at a $\Delta=\pm\omega_m$ detuning from the natural transition frequency whose amplitudes were determined by the driving strength. As the mechanical drive strength increased, the carrier was suppressed and the sideband amplitudes increased linearly with the square root of the power applied to the IDT. 

Coherent optomechanical control was then demonstrated in the time domain by performing Rabi oscillations on the first red sideband. The NV was initialized into the ground state with a 532 nm laser pulse then irradiated with a short 637 nm laser pulse on resonance with the first red sideband. The observation of Rabi oscillations demonstrated the coherent nature of the optomechanical control. The dependence of the Rabi frequency was shown to be linear in the square root of the IDT power, which agreed with the theoretical prediction. 

\section{Toward coherent interactions between defects and phonons}

\subsection{Phonon cooling}

The utility of future hybrid mechanical devices hinges on the ability to cool the mechanical degree of freedom to or near its quantum ground state. For instance, in devices where phonons serve as a flying qubit, the mechanical oscillator must be in a quantum coherent state. To prepare such a state, a two-level system can be used to cool the oscillator to its ground state, and subsequently prepare an arbitrary superposition of Fock states or other non-Gaussian states. This would also be of fundamental interest, as creation of such states cannot be accomplished with standard cavity optomechanical devices. On the other hand, as we show in the following section, phonon-mediated spin-spin entanglement can be performed for an oscillator in a thermal state. The entanglement fidelity is optimized when the oscillator is near its ground state. Achieving ground state cooling with a two-level system would provide a new approach to study cavity quantum electrodynamics using phonons.

There are now several theoretical proposals for cooling macroscopic mechanical oscillators with solid state two-level systems that draw heavily from the fields of cavity optomechanics \cite{HybridMechBook,AspelmeyerReview} and ultracold atomic physics\cite{MonroeCooling}. The general starting point for most cooling techniques requires operation in the resolved-sideband regime, in which the mechanical frequency exceeds the linewidth of the qubit transition, $\omega_m\gg\Gamma$. In this limit, a motional sideband of the carrier qubit transition can be selectively addressed, allowing for controlled interactions between the two-level system (TLS) and the mechanical oscillator \cite{CoolingReviewMonroe}. In particular, addressing the first red-detuned mechanical sideband results in an effective Jaynes-Cummings interaction, $H_{rsb}\sim\sigma^+a+\sigma^-a^{\dag}$, in which an excitation of the TLS results in a removal of a single phonon from the mechanical oscillator. Conversely, addressing the first blue-detuned mechanical sideband results in the anti-Jaynes-Cummings interaction, $H_{bsb}\sim\sigma^+a^{\dag}+\sigma^-a$,and an electronic excitation results in an addition of a single phonon to the mechanical oscillator. 

Interestingly, the electronic structure of the NV center allows for an additional cooling protocol which does not require addressing motional sidebands. The NV electronic structure facilitates a natural Jaynes-Cummings interaction between the orbital and spin degrees of freedom and phonons that does not require an external driving field. This attractive feature allows for resonant enhancement of the NV-phonon interaction and more efficient cooling. In this section we will outline the theoretical phonon cooling protocols established for both strain and magnetically mediated interactions. In the following section, we will evaluate the feasibility and challenges for accomplishing these protocols.

The TLS can be formed from the spin levels in the ground state manifold or from the orbital states of the NV center, with each offering distinct advantages for phonon cooling. For instance, cooling based on the ground state spin levels is more robust against external noise, and hence the spin-phonon interaction can persist for long times \cite{RablPRB,RablCoolingHighTemp}. In contrast, cooling using the orbital states benefits from the high sensitivity to mechanical motion and offers larger cooling powers \cite{Kepesidis}. As we will discuss, the use of stimulated Raman transitions between the ground state spin levels via virtual excitation of the orbital excited state can offer a cooling protocol that in some measure combines the coherence of the spin states with the strong strain-orbit coupling of the orbital excited state. These three cooling protocols are illustrated in Fig. 9.

\begin{figure}[!]
\includegraphics[width=8.6cm]{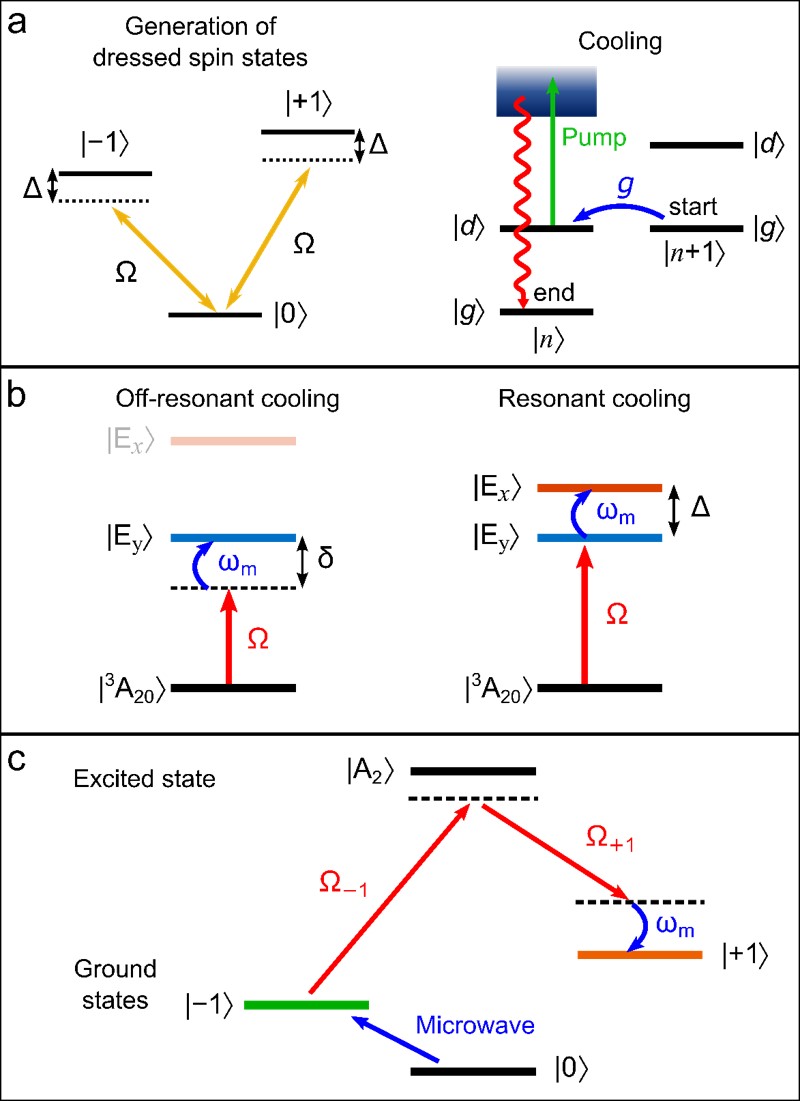}
\caption{Three cooling protocols for NV-mechanical oscillator systems. (a) Phonon cooling using the ground state spin. To generate resonant spin-phonon interactions, two external, near-resonant microwave fields address the $\ket{0}\leftrightarrow\ket{\pm 1}$ transitions with Rabi frequency $\Omega$ and detuning $\Delta$ and generate dressed spin states $\ket{g},\ \ket{d},$ and $\ket{e}$. Vibrational quanta are removed from the oscillator via spin excitations via the $\ket{g}\ket{n}\rightarrow\ket{d}\ket{n-1}$ transition. Subsequent optical pumping cools the resonator with negligible heating. (b) Phonon cooling using the orbital excited state. In the off-resonant cooling scheme, a laser is detuned from the $\ket{A}\rightarrow\ket{E_y}$ transition by $\delta=-\omega_m$, resulting in a removal of a single phonon from the mechanical oscillator. The NV releases this energy via spontaneous emission. In the resonant cooling scheme, the energy difference between $\ket{E_x}$ and $\ket{E_y}$, $\Delta$, is set to the mechanical frequency. The NV is prepared in the $\ket{E_y}$ state, and subsequently excited to $\ket{E_x}$ via removal of a phonon from the mechanical oscillator. (c) Cooling via stimulated Raman transitions. In this scheme, the red phonon sideband of the $\ket{\pm 1}\rightarrow\ket{A_2}$ transition is far off-resonantly excited with two laser beams with Rabi frequencies $\Omega_{\pm 1}$, and serves as the virtually excited level in the $\Lambda$ configuration. When the beatnote between the two lasers is equal to the qubit splitting, a Raman transition between the $\ket{\pm 1}$ spin levels occurs and allows for cooling in the $\ket{-1}\rightarrow\ket{1}$ transition.}
\end{figure}

The first cooling scheme, shown in Fig. 9a, utilizes a combination of optical pumping and resonant spin-phonon interactions \cite{RablPRB}. First, an excitation of the spin removes a single phonon from the mechanical oscillator. Through subsequent optical pumping, the spin is re-initialized and the phonon energy is dissipated. Typically, the spin-phonon interaction is parametric in nature, and hence a resonant exchange of vibrational and spin quanta requires an external, resonant driving field to prepare new dressed spin states and convert the parametric interaction into a transverse interaction. We note that the transverse spin-strain interaction naturally provides a transverse interaction without the need for external driving. This approach has the advantage of mitigating possible heating induced by the driving fields, but suffers from the inability to easily switch off the interaction, which may induce heating by resonant exchange. Here, we will limit the discussion to the parametric interaction, which was first proposed by refs. \cite{RablNatPhys,RablPRB}. 

To create the required dressed spin levels, near-resonant microwave fields can be applied at the $\ket{0}\leftrightarrow\ket{\pm 1}$ transition frequencies with Rabi frequency $\Omega$ and detuning $\Delta$. The resulting eigenstates are a dark state $\ket{d}=(\ket{1}-\ket{-1})/\sqrt{2}$ and two dressed states, $\ket{g}=\cos{(\theta)}\ket{0}-\sin{(\theta)}(\ket{1}+\ket{-1})/\sqrt{2}$ and $\ket{e}=\sin{(\theta)}\ket{0}+\cos{(\theta)}(\ket{1}+\ket{-1})/\sqrt{2}$, where $\tan{(\theta)}=-\sqrt{2}\Omega/\Delta$. If $\omega_{\ket{g}\leftrightarrow\ket{d}}=\omega_m$, a Jaynes-Cummings type spin-phonon interaction is formed with the interaction Hamiltonian $H_{int}=g^s(\ket{d}\bra{g}a+\ket{g}\bra{d}a^{\dag})$, where $g^s=\gamma_{NV}\frac{\partial B}{\partial z}z_0$ for magnetic coupling and $g^s=d_{\parallel}\epsilon_{\parallel}$ for axial strain coupling. 

The cooling protocol begins with an optical laser pulse that pumps the spin into $\ket{g}$. The spin-phonon interaction then allows for a spin transition, $\ket{g}\ket{n}\rightarrow\ket{d}\ket{n-1}$, that cools the mechanical mode. After this cycle occurs, a second laser pulse initializes the system to $\ket{g}$ without heating the resonator, and this cycle is repeated. With an optimal choice of $\Delta$, ground state cooling can be used to cool MHz frequency mechanical modes using magnetic coupling if the initial temperature of the system is low ($T\sim0.1-1$ K)\cite{RablNatPhys,shimonscience,RablPRB,RablCoolingHighTemp}.

The second cooling protocol relies on the orbital degree of freedom of the NV center and its strong coupling to crystal strain induced by mechanical motion (Fig. 9b). This type of protocol was first proposed by ref. \cite{WilsonRae}, and involved a system composed of a quantum dot that is parametrically coupled to a mechanical oscillator via strain. This technique was very recently adapted for the NV center by ref. \cite{Kepesidis}. The protocol relies on a laser-assisted electron-phonon interaction which converts the parametric interaction into an effective transverse interaction. The overall Hamiltonian for this system is described by a two-level system with ground state $\ket{g}$ and excited state $\ket{e}$, a harmonic oscillator, the parametric interaction, and a laser interaction term:

\begin{equation}
H=\frac{\omega_0}{2}\sigma_z+\omega_ma^{\dag}a+g(a+a^{\dag})\sigma_z+\Omega\cos{(\omega_Lt)}\sigma_x
\end{equation}

where $\sigma_z=\ket{e}\bra{e}-\ket{g}\bra{g}$, $\Omega$ is the optical Rabi frequency and $\omega_L$ is the laser frequency. For the NV center, it is desirable to use the $m_s=0$ spin sublevel for $\ket{g}$ and the $\ket{E_y}$ level for $\ket{e}$ due to their large response to crystal strain. When the laser is tuned to the first red sideband, $\omega_L=\omega_0-\omega_m$, cooling occurs with with a rate given by

\begin{equation}
\Gamma_c=\frac{g^2\Omega^2}{\Gamma\omega_m^2}
\end{equation}

where $\Gamma$ is the linewidth of the electronic transition. The corresponding rate equation describing the mean phonon number of the mechanical mode obeys

\begin{equation}
\frac{\partial\langle a^{\dag}a\rangle}{\partial t}=-\Gamma_c(\langle a^{\dag}a\rangle-n_f)
\end{equation}

where $n_f\approx\frac{\omega_m\bar{N}}{\Gamma_cQ}$ is the final thermal occupation number and $\bar{N}$ refers to the average phonon occupation of the thermal bath. In the limit where the cooling rate exceeds the thermalization rate of the mechanical mode, ground state cooling is possible. We note that this cooling scheme is valid in the Lamb-Dicke limit, where $\frac{g}{\omega_m}\sqrt{2\bar{n}+1}\ll\ 1$ and $\bar{n}$ is the mean thermal occupation number of the resonator.

This cooling protocol, sometimes referred to as ``off-resonant'' cooling, has an important disadvantage. The cooling rate scales inversely to the square of the phonon frequency which is a result of the fact that the transition is being addressed on a sideband and is natively off-resonant. This is complicated by the fact that generation of NV-phonon couplings large enough to induce cooling require nanoscale, high frequency mechanical resonators. To address this issue, another cooling protocol was proposed in ref. \cite{Kepesidis} that exploits the orbital-doublet structure of the NV excited state and allows for resonant cooling of the mechanical mode.

As described in section II. D., $E_2$ symmetric strain directly couples the $\ket{E_y}$ and $\ket{E_x}$ levels of the $^3E$ excited state. Therefore, when their frequency difference $\Delta_{xy}$ is tuned to the phonon frequency, a resonant exchange of energy can occur between the NV center and the mechanical resonator. The Hamiltonian in the interaction picture and within the rotating wave approximation can be written as

\begin{equation}
H_{int}=g_{E_2}\Big(\ket{E_x}\bra{E_y}a+\ket{E_y}\bra{E_x}a^{\dag}\Big)
\end{equation}

When the NV center is prepared in $\ket{E_y}$, it can be excited to the $E_x$ state by removing a single vibrational quantum from the mechanical oscillator. With spontaneous emission and constant driving the $\ket{0}\leftrightarrow\ket{E_y}$ transition with resonant laser light, the NV can efficiently cool the mechanical mode of interest as depicted in Fig. 9. The cooling rate associated with this process is given by

\begin{equation}
\Gamma_c=4\frac{g_{E_2}^2\Omega^2}{\Gamma^3}
\end{equation}

As discussed, the resonant cooling scheme does not depend on the phonon frequency. An important note is that this resonant cooling scheme should be more efficient from a practical standpoint. Given that the transverse and parametric NV-phonon couplings are typically of the same order of magnitude, that means significantly more laser power will be required in the off-resonant scheme to generate the same effective cooling power. Therefore, the resonant cooling scheme will be inherently less susceptible to heating from laser absorption.

The state of the mechanical resonator can be directly probed through excitation spectroscopy of the NV center. In the off-resonant coupling scheme, sideband thermometry can be used to extract the mean phonon number in the mechanical mode \cite{MonroeCooling,Kepesidis}. For the resonant cooling scheme, the mode temperature can be probed by measuring the intensity of $x$-polarized photons scattered from the NV center, which is directly proportional to the population in $\ket{E_x}$ and the mean thermal occupation number of the resonator. 

To achieve ground state cooling of a diamond mechanical oscillator, improvements on current devices must be made. So far, only the device presented in ref. \cite{Kenny} has met the baseline requirement of demonstrating strain-orbit coupling in a passive mechanical resonator. In that device, the single phonon coupling parameters were measured to be $g_{A_1}/2\pi\approx\ 1$ kHz, $g_{E_1}/2\pi\approx\ 3$ kHz, and $g_{E_2}/2\pi=0$, with an optical linewidth $\Gamma/2\pi=1$ GHz, phonon frequency $\omega_m/2\pi\approx\ 1$ MHz and quality factor $Q=20,000$. By scaling the device dimensions down to the nanoscale and improving optical linewidths, it should be possible to achieve ground state cooling with both the off-resonant and resonant cooling schemes. For instance, the fundamental flexural mode of a doubly-clamped diamond beam of dimensions 2 $\mu$m$\times$100 nm$\times$50 nm with $\omega_m/2\pi=230$ MHz and $Q=10^5$ that is coupled to a shallow NV center with a narrow optical linewidth of $\Gamma=100$ MHz would generate a sizable parametric single phonon coupling $g=21.5$ MHz. If the device is operated at a bath temperature of 4 K, then ground state cooling can be achieved with a single NV center using the off-resonant coupling scheme \cite{Kenny}. Interestingly, this device would also reside in the high cooperativity regime, with $C\sim 5$. This would allow for preparation and observation of non-Gaussian quantum states of the mechanical oscillator \cite{MonroeCatStates,MonroeNonClassical,cleland2010}.

The final cooling protocol is based on stimulated Raman transitions between two ground state spin levels via the orbital excited state of the NV center \cite{Habraken,RamanNanodiamond}. In this $\Lambda$ configuration, both the long coherence times of the spin and the strong phonon coupling of the orbitals can be accessed simultaneously. The Raman cooling process is shown schematically in Fig. 9c. For illustrative purposes, we consider Raman transitions coupling the $\ket{\pm 1}$ spin levels via the $\ket{A_2}$ excited state. In this situation, the parametric strain-orbit interaction creates phonon sidebands on the $\ket{\pm 1}\leftrightarrow\ket{A_2}$ transitions, as described above. Because the Lamb-Dicke parameters for the $\ket{\pm 1}\leftrightarrow\ket{A_2}$ transitions are the same, the Raman transition must involve both an optical carrier transition and an optical red sideband transition in order to get an effective spin-phonon coupling \cite{RamanNanodiamond}. Hence, Raman cooling requires operation in the resolved-sideband regime in the optical domain. Moreover, to adiabatically eliminate the excited state, the Raman beams must be far detuned from the sideband transition ($\Delta_{\pm 1}\gg \Omega_{\pm 1}$), while remaining spectrally isolated from the carrier transition or other sidebands. 

The cooling process begins by initializing the NV into $\ket{-1}$ and applying two lasers of frequencies $\omega_{-1}$ and $\omega_1$ to the NV center that are $\sigma^-$ and $\sigma^+$ polarized and generate optical Rabi frequencies $\Omega_{-1}$ and $\Omega_1$. When the two-photon resonance condition $\omega_1-\omega_{-1}=\omega_{\pm 1}-\omega_m$ is met and the detunings from the excited state are large compared to the single photon Rabi frequencies, then an effective spin-phonon interaction arises of the form

 \begin{equation}
 H_{eff}=\frac{\eta\Omega_{-1}\Omega_1}{\Delta}(\ket{1}\bra{-1}a+\ket{-1}\bra{1}a^{\dag})
 \end{equation} 
 
 where $\omega_{\pm 1}$ is the qubit splitting and we assume $\Delta=\Delta_{-1}=\Delta_{1}$ for sufficiently large detunings. Accompanying this spin-phonon interaction is an additional decay channel through spontaneous emission, which is quantified by $\Gamma_{eff}=\Gamma\frac{\Omega_{\pm 1}^2}{\Delta^2}$. 
 
 A Raman transition $\ket{-1}\rightarrow\ket{1}$ removes a single phonon from the mechanical oscillator. From there, a double quantum magnetic pulse can be used to re-initialize the spin into $\ket{-1}$ for more cooling. While this cooling protocol requires more overhead, it has the advantage of tuning the effective spin-phonon coupling and decay rates by changing the laser frequency and power. Moreover, it is directly compatible with existing protocols for phonon-mediated spin-spin entanglement \cite{RamanNanodiamond}.
 
 \begin{figure*}[!]
\includegraphics[width=14cm]{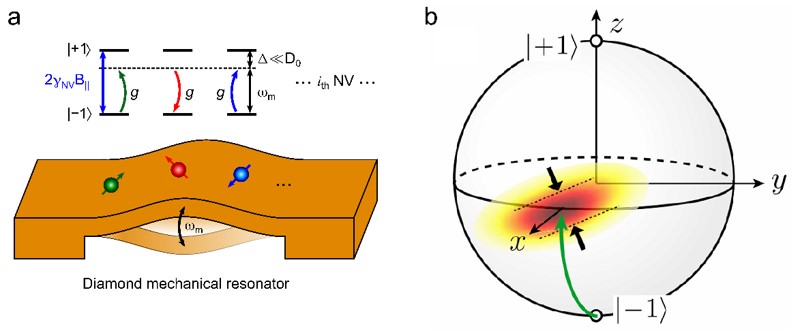}
\caption{Phonon-mediated spin-spin interactions. (a) Multiple NV centers embedded in a diamond mechanical resonator couple to each other through a common mechanical mode. Because the phonon extends over the entire structure, long-range interactions between spins are possible. (b) Dispersive coupling of an ensemble of NV spins to a common mechanical mode allows for squeezing of the spin uncertainty distribution. Here, the $x$ quadrature of the spin variance is squeezed as discussed in the main text. Figure reprinted with permission from \cite{Bennett}. Copyright Physical Review Letters}
\end{figure*}

\subsection{Phonon mediated spin-spin interactions}

An outstanding challenge in quantum information science is the generation of controlled interactions between qubits. To date, NV centers have been entangled through dipolar interactions \cite{DipolarEntanglement} and photon-mediated interactions \cite{BernienEntanglement,LoopholeFree}. The dipolar coupling between two spins separated by a distance $d$ scales as $d^{-3}$, and thus entanglement typically requires NVs to be within a few tens of nanometers from each other, which is infeasible to achieve in a controlled manner with current capabilities. Photon-mediated entanglement has been used to entangle NV centers over large distances and with high fidelity, but can be technically challenging due to the probabilistic, heralded nature of entanglement and the requirement of indistinguishable photons \cite{BernienEntanglement,LoopholeFree,UnconditionalTeleportation,AlpPRL,BernienPRL}. Phonon-mediated interactions offer a promising alternative to these schemes by allowing for deterministic, long-range interactions between NV spins. Generation of entanglement requires operation in the large cooperativity regime \cite{Bennett,RablNatPhys}, where $C=\frac{g^2}{\Gamma_2\gamma}>1$, and $\Gamma_2$ is the spin dephasing rate and $\gamma$ is the phonon thermalization rate.

Phonon-mediated spin-spin interactions have been proposed in a variety of hybrid architectures, utilizing magnetic fields \cite{RablNatPhys,RablPRB} and strain \cite{Bennett,GeometricPhase,RamanNanodiamond,DeterministicEntanglementDu,DistantEntanglementPRA,LumingNanodiamond,EntanglementMagneticFilm} as an a coupling interface. While their executions may differ experimentally, their physical foundations are the same. As an illustrative example, we consider two NV center spins that are strain-coupled to a nanomechanical resonator. Specifically, we focus on the $\ket{\pm 1}$ spin levels as a qubit and the $E$ symmetric spin-phonon interaction, which directly couples the $\ket{\pm 1}$ spin levels. The Hamiltonian describing the hybrid system in the rotating wave approximation is given by \cite{Bennett}

\begin{equation}
H=\sum_{j=1}^2\left[\omega_j\sigma_j^+\sigma_j^-+g_j(\sigma_j^+a+\sigma_j^-a^{\dag})\right]+\omega_ma^{\dag}a
\label{eq:twoqubit}
\end{equation}

where $\omega_j$ is the energy of the $j$-th qubit, $\sigma_j^{+(-)}$ is the spin raising (lowering) operator for the $j$-th qubit, and $g_j$ is the single phonon coupling for the $j$-th qubit. Most proposals consider the dispersive coupling regime in which $g_j\ll\delta=\omega_j-\omega_m$. In this regime, spin-spin interactions are mediated by a virtual exchange of phonons. To obtain the effective spin-spin interaction, we move into the interaction picture with respect to the qubit and oscillator and apply the unitary transformation $UHU^{\dag}$, where the unitary operator $U$ is given by

\begin{equation}
U=\text{exp}\left[\sum_{j=1}^2\frac{g_j}{\delta_j}(\sigma_j^+a-\sigma_j^-a^{\dag})\right]
\end{equation}

For simplicity, the detunings $\delta_j$ and single phonon couplings $g_j$ are assumed to be equal and are defined as $g$ and $\delta$. In the dispersive regime, the unitary transformation can be carried out to second order in $(g/\delta)$, and the spin-spin interaction Hamiltonian is given by

\begin{equation}
\begin{split}
H_{eff}=&\left(\delta+\frac{2g^2}{\delta}(1+a^{\dag}a)\right)(\sigma_1^+\sigma_1^-+\sigma_2^+\sigma_2^-)\\
&+\frac{g^2}{\delta}(\sigma_1^+\sigma_2^-+\sigma_1^-\sigma_2^+)
\end{split}
\end{equation}

In the first term, we see an effective AC Stark shift due to the dispersive coupling and in the second term we have an effective spin-spin interaction. By choosing a suitable gate time, the spins can be completely disentangled from the mechanical resonator and entangled with each other. For instance, proper gate timing and an initial preparation of both qubits in an eigenstate of $\sigma_{\phi}=\cos{(\phi)}\sigma_x+\sin{(\phi)}\sigma_y$ leads to the two-qubit entangled state $\ket{\psi}=\frac{1}{2}(\ket{1}_1\ket{-1}_2+e^{i\phi}\ket{-1}_1\ket{1}_2)$. When integrated into a phonon router, this entangling scheme can be used to deterministically shuttle information between distant NV centers where phonons serve as a flying qubit \cite{Habraken,UniversalSAW,Fang}.

As mentioned, these applications require operation in the high cooperativity regime. To date, no such devices have been demonstrated in this regime with NV centers as shown in Table 1. For magnetically coupled devices, it should be possible to enter the strong coupling regime by designing nanoscale resonators with large zero-point motion amplitudes and employing dynamical decoupling. For instance, for the magnetized cantilever tip device scheme described in ref. \cite{shimonscience}, designing a resonator with $Q>10^6$, $\omega_m/2\pi\sim 1$ MHz, and a field gradient $\frac{\partial B}{\partial z}\sim 10^5$T/m and coupling it to a shallow spin with a coherence time on the millisecond time scale, it should be possible to enter the high cooperativity regime. In the case of strain-coupled devices, it should be possible to enter the high cooperativity regime using both the spin and orbital degrees of freedom. For instance, the nanobeam described in the previous section could also achieve the high cooperativity regime for spin-strain coupling by operating at dilution refrigerator temperatures ($T\sim100$ mk) \cite{NatureComm}. In the final section, we will evaluate the challenges facing the development of these types of devices and offer promising alternative architectures for hybrid NV-mechanical devices.

In addition to entanglement, this scheme can be use to generate spin-squeezed states in large ensembles of NV centers (Fig. 10). If the total angular momentum of the spin ensemble is conserved, the effective Hamiltonian for this device can be written as \cite{Bennett}

\begin{equation}
H_{eff}=\left(\delta+\frac{2g^2}{\delta}(1+a^{\dag}a)\right)J_z+\frac{g^2}{\delta}(\mathbf{J}^2-J_z^2)
\end{equation}

where $\mathbf{J}$ is the collective spin operator. The term containing $J_z^2$ corresponds to a one-axis twisting Hamiltonian which can squeeze the spin uncertainty distribution in a single direction. If prepared in an eigenstate of $J_x$, the spins precess around the equator of the collective Bloch sphere with a precession rate proportional to $J_z$, which shears the uncertainty distribution and hence reduces the spin variance in the $x$ direction, as shown in Fig. 10 b. The squeezing is quantified by the so-called squeezing parameter, $\xi$, which is defined to be

\begin{equation}
\xi^2 = \frac{2J\langle\Delta J_{min}^2\rangle}{\langle J_x\rangle^2}
\end{equation}

where $\langle\Delta J_{min}^2\rangle$ quantifies the minimum spin uncertainty. A spin squeezed state is quantified by $\xi^2<1$, and would enable magnetometry with a precision below the quantum projection noise limit. Bennett $et\ al.$ showed that spin squeezing can occur for an ensemble of $N=100$ NV spins under realistic experimental conditions, including spin dephasing and phonon number fluctuations. The phonon-induced A.C. Stark shift is proportional to $a^{\dag}a$ and hence phonon number fluctuations will induce additional dephasing to the typical inhomogeneous dephasing of NV ensembles. By using a combination of spin-echo pulse sequences and mechanical driving, it is possible to mitigate these sources of decoherence and achieve an optimal spin squeezing of

\begin{equation}
\xi_{opt}\simeq \frac{2}{\sqrt{JC}}
\end{equation}

where $J$ is the total angular momentum of the spin ensemble and $C$ is the cooperativity. Note that spin squeezing does not require operation in the high single-phonon cooperativity regime. 

\section{Prospects and challenges for future devices}

In this section, we will evaluate the prospects and challenges for future hybrid mechanical devices utilizing the NV center. In particular, we will discuss the challenges of reaching the high cooperativity regime and discuss realistic architectures for new devices that can realize quantum coherent interactions between the NV center and a mechanical oscillator. 

The single phonon cooperativity parameter can be written in a more useful way that is explicit in terms of experimentally relevant parameters. 

\begin{equation}
C=\frac{2\pi g^2QT_2}{\bar{n}\omega_m}\sim\frac{h g^2QT_2}{k_BT}
\end{equation}

The recipe for residing in the high cooperativity regime is clear: increase the single phonon coupling, increase the mechanical quality factor,  reduce the system temperature, and increase the coherence time of the NV center. However, given current experimental limitations, it is difficult to achieve all of these tasks simultaneously. For magnetically coupled devices, the single phonon coupling strength is primarily limited by the magnetic field gradient \cite{shimonscience,ArcizetNatPhys}. For strain coupled devices, the coupling strengths are limited by the strain induced by the zero point motion of the resonators, which scales inversely with the dimensions of the resonator \cite{NatureComm,Teissier,Meesala} and hence proportionally to the mechanical frequency. In both cases, these couplings can be increased by bringing the target NV center closer to the surface of diamond.  For magnetic coupling, the NV should be as close as possible to the field gradient source, and hence as close to the diamond surface as possible. For strain coupling, the NV should be close to the surface of diamond, where the strain is typically largest, and moreover, by reducing the dimensions of the resonator to increase the zero-point strain energy, the NV will naturally be closer to diamond surfaces. At odds with this requirement is the fact that the coherence properties of the NV center are significantly degraded when brought into close proximity of the diamond surface \cite{BryanPRL,RugarEField,RosskopfPRL,DegenPRB,SurfaceNoiseBarGill}. In particular, magnetic field noise due to a surface bath of rapidly flipping electronic spins and electric field noise due to surface dipoles or lattice vacancies reduce the spin coherence times of NV centers located within 25 nm of the diamond surface by one to three orders of magnitude \cite{BryanPRL}. In addition, spectral diffusion due to a combination of electric field noise and NV charge state fluctuations can induce orbital dephasing and broaden the optical transition linewidths of the NV center by several orders of magnitude.

To date, the longest spin coherence times recorded for NV centers have been for NVs located deep in bulk diamond samples. Under ambient conditions, the longest $T_2$ time ever recorded is on the order of milliseconds \cite{IsotopeEngineering}. At liquid nitrogen temperatures, $T_2$ times on the order of 500 ms were observed \cite{BarGillNatComm}, and required many refocusing, spin-echo pulses. In addition, narrow optical linewidths of $\Gamma=17$ MHz were observed in a single spectroscopic measurement on an NV center in a diamond nanocrystal \cite{BestLineNanocrystal}. These linewidths are close to the lifetime limited linewidth, $\Gamma=13$ MHz, that was observed for NVs in a bulk diamond sample \cite{BestLineEver}. These results suggest that mitigating surface noise should allow for millisecond spin coherence times and optical linewidths of tens of MHz \cite{BassettPRL,AcostaPRL,YellowLaser,BestLineNanocrystal,AslamNJOP}.
 
Correspondingly, a significant effort in the diamond community is being made to address surface-induced spin decoherence and spectral diffusion through a variety of techniques, including surface passivation \cite{ShanyingFluorine}, downstream etching \cite{ReducedDamage}, high temperature annealing\cite{HighTAnneal}, and alternative NV formation techniques \cite{DeltaDoping,OhnoAPL,Claire,OhnoCImplantation,HeliumImplantation}. 
NV centers naturally occur in diamond, but can be artificially introduced through nitrogen ion implantation, which is the most common approach taken to introduce NV centers. In general, the spin coherence properties of these implanted NV centers are inferior to naturally occuring NV centers. This degradation in spin coherence is attributed to lattice damage introduced during the implantation process, poor nitrogen to NV center conversion efficiencies, and plasma-induced damage during etching \cite{ReducedDamage}. In 2012, two groups \cite{OhnoAPL,DeltaDoping} introduced a nitrogen delta-doping technique to form NV centers that is inherently gentler than ion implantation and mitigates lattice damage. In this method, nitrogen gas is introduced during CVD growth of diamond for a short amount of time, introducing a nitrogen-rich layer into the diamond sample. The sample can then be electron irradiated to create vacancies and annealed to create NV centers. This process has been successful in creating shallow NV centers with long coherence times, with an observed $T_2>200\ \mu$s for a 20 nm deep NV center \cite{Claire} and an observed $T_2>100\ \mu$s for a 5 nm deep NV center \cite{OhnoAPL}.

\begin{figure*}[!]
\includegraphics[width=17.8cm]{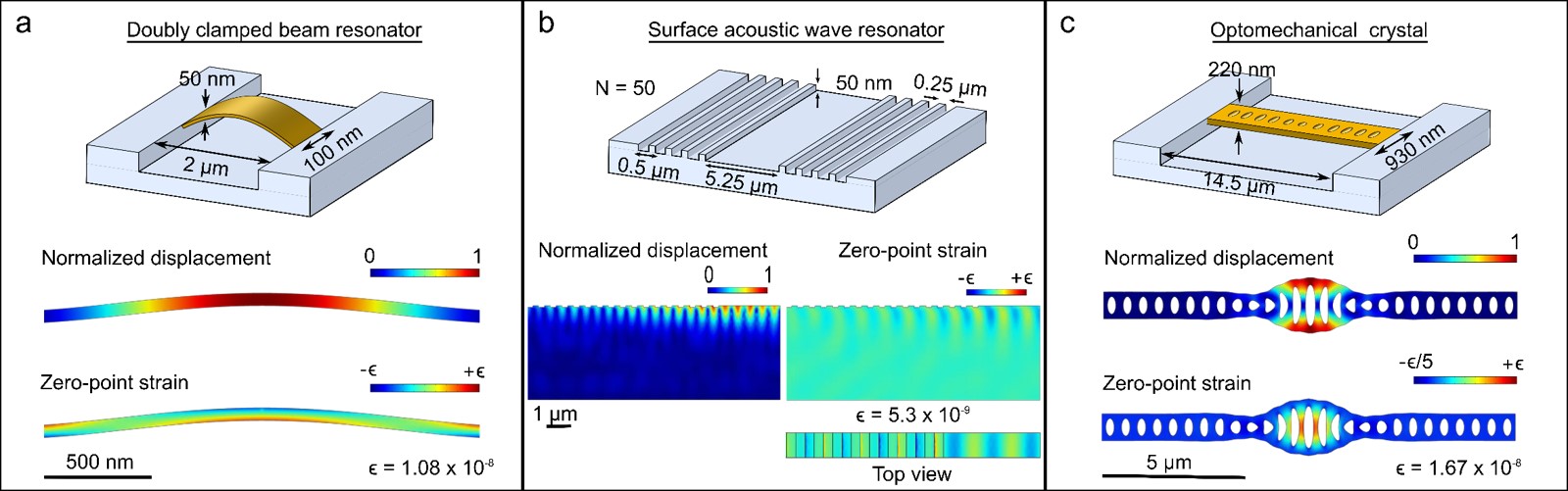}
\caption{Example devices with potential to operate in the high cooperativity regime. (a) A doubly-clamped nanobeam with dimensions 2 $\mu$m$\times$100 nm$\times$50 nm and $\omega_m/2\pi=230$ MHz.(b) SAW cavity with a confined phonon mode $\omega_m\sim$ 10 GHz defined by the grating pitch and the speed of sound of diamond. (c)  Optomechanical crystal with dimensions $14.5\ \mu m\times\ 930$ nm$\times\ 220$ nm and a localized, mechanical breathing mode with $\omega_m/2\pi\sim$6 GHz. For each device, simulations of the mode displacement and zero point strain are displayed underneath the device schematic. These are calculated with finite element simulations.}
\end{figure*}

To mitigate lattice damage, which is presumed to be partially responsible for spectral diffusion of the zero-phonon line, the diamond sample can be annealed at high temperatures (1200$^{\circ}$C). At these temperatures, vacancies can move to the surface and be annealed away, essentially repairing the diamond lattice. This technique was shown to produce narrow optical lines in an implanted diamond sample \cite{HighTAnneal}, with the best observed linewidth of $\Gamma=27$MHz. Moreover, this technique was used to improve the spin coherence times of implanted NV centers and stabilize the NV charge state \cite{HighTAnnealSpins}. These materials science approaches are promising for engineering shallow NV centers with excellent coherence and optical properties.

In terms of the mechanics, single-crystal diamond should in principle be an excellent substrate material for high quality factor mechanical resonators due to its low intrinsic losses. Indeed, mechanical quality factors exceeding one million have already been demonstrated in SCD cantilevers at cryogenic temperatures \cite{DegenNatComm}. However, as the mechanical frequency increases to improve the single phonon couplings, the quality factor correspondingly decreases while the so called $Q\cdot f$ product remains roughly constant. The reduction in quality factor can be attributed to a number of sources, including two-level defects, surface chemistries, surface roughness,and clamping losses\cite{DegenNatComm,PreetiAPL,BurekAPL}. However, achieving high $Q$ diamond mechanical oscillators with high frequencies is not an impossible task. For instance, the clamped beams that are typically used in these experiments can be designed to avoid clamping losses \cite{GarrettCole}, and with the continual improvements and advancements in diamond processing, these devices could operate in the high cooperativity regime. In addition, a phononic shield placed around the device could also significantly improve the quality factor \cite{PainterNature}.

The challenge of increasing $g$ while mitigating dissipation and decoherence may be overcome by choosing new types of architectures for hybrid NV-mechanical devices. In particular, diamond optomechanical crystals and surface acoustic wave cavities currently stand out as promising avenues for reaching the high cooperativity regime. SAW cavities have already been used to generate GHz frequency phonon modes in other substrate materials with high quality factors approaching one million at low temperatures \cite{HighQSAW1,HighQSAW2,QuantumSAW}, and more importantly, they have been demonstrated in the quantum regime \cite{Gustafsson,GustafssonNatPhys,QuantumSAW}. Furthermore, SAW phonons propagate approximately one acoustic wavelength into the surface of the elastic material. Because diamond has a high speed of sound, GHz frequency phonons would propagate $\sim\ 1 \mu$m into the diamond. This would allow the use of naturally occurring, deep NV centers, which have exhibited the best coherence and optical properties to date. We note that ref. \cite{Golter} has already utilized this benefit in experiments to use SAWs to perform optomechanical control of naturally occurring NV centers. However, because there was no cavity element in the device, it cannot be extended to the quantum regime of coupling. SAW cavities in principle should also increase the coupling strength. Finite element simulations performed in COMSOL (see Fig. 11b) reveal that for a SAW cavity device supporting a 10 GHz phonon mode, $g_{A_1}=10$ MHz can be achieved. Assuming a reasonable $Q=10^5$ and optical linewidth $\Gamma=100$ MHz, it should be possible to enter the high cooperativity regime with the orbital states of the NV center.

Optomechanical crystal devices provide a promising platform for NV-mechanical experiments due to their potential to generate high single phonon couplings, isolate the NV center from surface noise, and mitigate clamping losses. Optomechanical crystals are structures that support both optical and mechanical resonances. More specifically, they are designed in such a way that the mechanical and optical modes of interest are coupled to each other, and are defect states that lie within a quasi-phononic and quasi-photonic bandgap respectively \cite{OptomechanicalCrystals}. The coupling allows for radiation pressure forces to actuate the mechanical mode. The partial phononic bandgap prevents phonons of the defect mode from tunneling out of the mechanical structure, improving the $Q$ \cite{PainterNature}. Moreover, this bandgap can be designed to increase the Q of a high frequency mechanical mode whose strain profile is constant through the cross section of the structure. For instance, consider the 1-D diamond optomechanical crystal shown in Fig. 11c, which has dimensions $14.5\ \mu m\times\ 930$ nm$\times\ 220$ nm. This crystal supports a localized, mechanical breathing mode, with an eigenfrequency of $\sim$ 6 GHz. The strain induced by the motion of this mode is highest at the center of the length of the beam. However, the strain is also constant through out the thickness of the beam. This means that a target NV center could be located at the volumetric center of this beam, with the nearest diamond surface being approximately 100 nm away. Since the NV would not be in close proximity to a surface, its coherence and optical properties should in principle resemble those in bulk diamond samples \cite{BryanPRL}. If the NVs are introduced using nitrogen delta-doping, then these coherence properties should not be significantly affected \cite{OhnoAPL,DeltaDoping}. Therefore, a ground state spin coherence time $T_2^s=10$ ms and orbital coherence time $T_2^o=10$ ns should be reasonably attainable in this device \cite{OhnoAPL,HighTAnneal}. Moreover, a quality factor of $Q=10^4$ should also be attainable for this mechanical mode, which is consistent with recent measurements of diamond optomechanical crystals \cite{DiamondOMCrystals}.

Diamond optomechanical crystals can not only offer improvements to both the mechanical and NV decoherence, but also offer improved NV-phonon couplings. For the mechanical mode depicted in Fig. 11c, the orbital single phonon couplings would be $g^o\sim$ 20 MHz, with the spin single phonon coupling being $g^s\sim$ 1 kHz. With these improved couplings, and assuming the reasonable parameters of $Q=10^4$, $T_2^s=10$ ms, $T_2^o=10$ ns, the single phonon cooperativities should be $C^o\sim 3$ for an operation temperature of 4 K and $C^s\sim 0.4$ for an operation temperature of 100 mK.

In addition to the NV center, the mechanical mode of the optomechanical crystal can directly couple to the photonic modes of the crystal via radiation pressure \cite{PainterNature}. Therefore, this device could achieve mechanical cooling that combines radiation pressure back action cooling with single defect cooling. This paves the way for studies of the interplay between the two interactions and cavity quantum electrodynamics (cQED) experiments in diamond optomechanical crystals \cite{Kimble,Restrepo}.

\section{Conclusion}
As discussed in this review, the development of high quality factor mechanical structures coupled with ever-improving diamond processing has enabled novel micromechanical and nanomechanical devices that can be used to control and manipulate NV centers in diamond. As the field advances toward the reverse process in which NV centers are used to control and manipulate mechanical states, several fundamental challenges must be overcome. Coherent interactions between the internal degrees of freedom of the NV center and phonons require that their intrinsic coupling exceeds the decoherence rates of the system. As we discussed, many efforts have been made to improve the coherence times and optical linewidths of single NV centers. Moreover, new mechanical devices are being pursued that should offer high frequency phonon modes with high quality factors. The ultimate challenge will be to develop a device that can accomplish both. As we showed, there are architectures for hybrid mechanical devices that should meet all of these requirements, and more importantly, be scalable to a large number of resonators and NV centers. The progress reviewed in this article shows that hybrid, quantum mechanical systems based in diamond are feasible and can be central components to the success of quantum information processing and studies of fundamental quantum phenomena.

\begin{acknowledgements}

This work is supported by the Air Force Office of Scientific Research Quantum Memories MURI program and an NSF-CAREER award (DMR-1352660). D. Lee acknowledges support by the KIST Institutional Program (Project No. 2E26681). 

\end{acknowledgements}

\bibliography{ReviewPaper_Final}

\end{document}